\documentclass[12pt,a4paper]{article}
\pdfoutput=1
\usepackage{jheppub}
\usepackage{epstopdf}
\usepackage{graphicx}
\usepackage{epsfig}
\usepackage{dcolumn}  
\usepackage{bm}    
\usepackage{amssymb} 
\usepackage{amsmath,bm}
\usepackage{amsfonts}    
\usepackage{slashed}  
\usepackage{youngtab}
\usepackage[mathscr]{euscript}
\usepackage{epsfig}
\usepackage{verbatim}
\newdimen\tableauside\tableauside=1.0ex
\newdimen\tableaurule\tableaurule=0.4pt
\newdimen\tableaustep
\def\phantomhrule#1{\hbox{\vbox to0pt{\hrule height\tableaurule width#1\vss}}}
\def\phantomvrule#1{\vbox{\hbox to0pt{\vrule width\tableaurule height#1\hss}}}
\def\sqr{\vbox{%
		\phantomhrule\tableaustep
		\hbox{\phantomvrule\tableaustep\kern\tableaustep\phantomvrule\tableaustep}%
		\hbox{\vbox{\phantomhrule\tableauside}\kern-\tableaurule}}}
\def\squares#1{\hbox{\count0=#1\noindent\loop\sqr
		\advance\count0 by-1 \ifnum\count0>0\repeat}}
\def\tableau#1{\vcenter{\offinterlineskip
		\tableaustep=\tableauside\advance\tableaustep by-\tableaurule
		\kern\normallineskip\hbox
		{\kern\normallineskip\vbox
			{\gettableau#1 0 }%
			\kern\normallineskip\kern\tableaurule}%
		\kern\normallineskip\kern\tableaurule}}
\def\gettableau#1 {\ifnum#1=0\let\next=\null\else
	\squares{#1}\let\next=\gettableau\fi\next}

\newcommand{\be}{ \begin{equation}}
\newcommand{\ee}{\end{equation}}
\newcommand{\bea}[1]{\begin{eqnarray}\label{#1} }
\newcommand{\eea}{\end{eqnarray}}

\def\ZZZ{{\hskip-3pt\hbox{ Z\kern-1.6mm Z}}}
\def\zzz{{\hskip-3pt\hbox{ z\kern-1mm z}}}

\def\s{\sigma}
\def\bal#1\eal{\begin{align}#1\end{align}}

\def\one{{\hbox{ 1\kern-.8mm l}}}
\def\zero{{\hbox{ 0\kern-1.5mm 0}}}

\def\>{\rangle}
\def\<{\langle}

\title{The supersymmetric affine Yangian}

\author{Matthias R.\ Gaberdiel$^a$,  Wei Li$^b$,  Cheng Peng$^c$, and Hong Zhang$^b$} 
\affiliation{$^a$ Institut f\"ur Theoretische Physik, ETH Zurich, \\
\hspace*{0.3cm}CH-8093 Z\"urich, Switzerland}
\affiliation{$^b$ Institute of Theoretical Physics, Chinese Academy of Sciences\\
\hspace*{0.3cm}100190 Beijing, P.R.\ China}
\affiliation{$^c$ Department of Physics, Brown University, \\
\hspace*{0.3cm}182 Hope Street, Providence  , RI 02912, USA}

\emailAdd{gaberdiel@itp.phys.ethz.ch, weili@itp.ac.cn, cheng$\underline{~}$peng@brown.edu, kilar@itp.ac.cn}

\abstract{The affine Yangian of $\mathfrak{gl}_1$ is known to be isomorphic to ${\cal W}_{1+\infty}$, the $W$-algebra that characterizes the bosonic higher spin -- CFT duality. In this paper we propose some of the defining relations of the Yangian that are relevant for the ${\cal N}=2$ superconformal version of ${\cal W}_{1+\infty}$. Our construction is based on the observation that the ${\cal N}=2$ superconformal ${\cal W}_{1+\infty}$ algebra contains two commuting bosonic ${\cal W}_{1+\infty}$ algebras, and that the additional generators transform in bi-minimal representations with respect to these two algebras. The corresponding affine Yangian can therefore be built up from two affine Yangians of $\mathfrak{gl}_1$ by adding in generators that transform appropriately.}

\begin{document}

\maketitle

\makeatletter
\g@addto@macro\bfseries{\boldmath}
\makeatother

\section{Introduction}

The tensionless limit of string theory on AdS is believed to be dual to a free (or nearly free) conformal field theory \cite{Sundborg:2000wp,Witten,Mikhailov:2002bp}, where string theory is expected to contain a higher spin theory \cite{Vasiliev:2003ev}. 
At this point in moduli space the large symmetry algebra underlying string theory (see \cite{Gross:1988ue,Witten:1988zd,Moore:1993qe} for indirect evidence) is expected to become visible. 
This is also the place where the integrability of the theory should be most easily discerned. 

In the context of AdS$_3$, the emergence of a higher spin symmetry at the tensionless point was recently seen quite explicitly in \cite{Gaberdiel:2014cha}, see also \cite{Gaberdiel:2017oqg,Ferreira:2017pgt} for attempts to observe this directly from a world-sheet perspective. 
In that case, the dual 2d conformal field theory of string theory on AdS$_3\times {\rm S}^3 \times \mathbb{T}^4$, the symmetric orbifold of $\mathbb{T}^4$, was shown to contain a ${\cal W}_\infty$ symmetry algebra. 
This is the hallmark of the duality between higher spin theories on AdS$_3$ and 2d CFT's \cite{Henneaux:2010xg,Campoleoni:2010zq,Gaberdiel:2010pz}, see \cite{Gaberdiel:2012uj} for a review. 

On the other hand, there has also been progress in understanding the integrable structure of string theory on AdS$_3$ \cite{OhlssonSax:2011ms,Sax:2012jv,Borsato:2015mma,Sfondrini:2014via}, and it would be very interesting to relate the higher spin and integrable symmetries. 
Integrable theories are usually distinguished by having a Yangian symmetry, and one may therefore try to identify the relevant Yangian in the explicit higher spin description. 
This was recently done \cite{Prochazka:2015deb,Gaberdiel:2017dbk} for the bosonic toy model of \cite{Gaberdiel:2010pz}, where the generators of ${\cal W}_\infty[\lambda]$, the symmetry algebra of the higher spin theory, were explicitly identified with those of the affine Yangian of $\mathfrak{gl}_{1}$. 
(The underlying isomorphism was first noted in \cite{Tsymbaliuk,Tsymbaliuk14}, generalizing the construction of \cite{GautamLaredo}, and independently by \cite{Miki} and \cite{FFJMM1,FFJMM2,FJMM1}, see also \cite{Kimura:2015rgi} for further generalizations. 
The affine Yangian of $\mathfrak{gl}_1$ is also isomorphic to the spherical degenerate double affine Hecke algebra ${\rm SH}^c$ of \cite{SV}, and was also constructed independently in \cite{MO}.)
\smallskip

In this paper we show how to construct the Yangian algebra corresponding to the ${\cal N}=2$ superconformal generalization of  ${\cal W}_\infty$. 
Our approach is partially inspired by the fact that the underlying higher spin algebra ${\rm shs}[\lambda]$ contains two commuting bosonic higher spin subalgebras ${\rm hs}[\lambda] \oplus {\rm hs}[1-\lambda]$. Subsequently, it was suggested in  \cite[section 11.1]{Gaiotto:2017euk} that this relation may also be true for the full ${\cal W}^{({\cal N}=2)}_\infty[\lambda]$ algebra. 
We begin by showing that ${\cal W}^{({\cal N}=2)}_\infty[\lambda]$ actually has this structure, i.e.\ that it contains two decoupled ${\cal W}_\infty[\mu]$ algebras. (This analysis relies on the precise form of the defining structure constant of the ${\cal W}^{({\cal N}=2)}_\infty[\lambda]$ algebra that was identified in \cite{Candu:2012tr}.) 
We then show that the additional generators that have to be added to the two bosonic ${\cal W}_\infty$ algebras in order to generate the full ${\cal W}^{({\cal N}=2)}_\infty[\lambda]$ algebra transform in what one may call bi-minimal representations with respect to the two ${\cal W}_\infty$ algebras. 
(This is to say, they transform as a minimal representation with respect to one, and as an anti-minimal representation with respect to the other; here ``anti-minimal" means that it is the conjugate representation to the minimal representation.) 
The basic idea of our construction is then to add generators to the two affine Yangians of $\mathfrak{gl}_1$ that have these transformation properties. 

The main technical difficulty of this approach comes from the fact that the description of conjugate minimal representations in terms of the affine Yangian was not known. 
The affine Yangian viewpoint gives rise to an elegant description of representations in terms of plane partitions \cite{Prochazka:2015deb},  see also \cite{Datta:2016cmw}, but this language only applies to the ``box"-representations, but not to those made of ``anti-boxes". However, the bi-minimal representations that are relevant for the above extension always involve also anti-box representations. We propose a general formula for the description of anti-box representations in terms of plane partitions, see Section~3.2. With this insight we can then propose some of the commutation relations of the two sets of affine Yangian generators of $\mathfrak{gl}_1$ with the additional modes, and thus undertake the first steps towards defining the supersymmetric generalization of the affine Yangian.  
\medskip

The paper is organized as follows. 
In Section~2 we show that the  ${\cal W}^{({\cal N}=2)}_\infty[\lambda]$ algebra contains (and can be built up from) two commuting bosonic ${\cal W}_\infty$ algebras. 
We identify the additional generators that need to be added, and in particular, their representation properties with respect to the two bosonic ${\cal W}_\infty$ algebras. 
In Section~3 we review the relevant minimal representation of the affine Yangian, and explain how to describe the conjugate representation. In Section~4 we analyze the ${\cal N}=2$ construction for $\lambda=0$, where the ${\cal W}^{({\cal N}=2)}_\infty[\lambda]$ algebra has a free field realization, in terms of which also the two bosonic ${\cal W}_\infty$ algebras can be identified. 
In particular, we can make an explicit ansatz for the additional generators that need to be added and compute their commutation and anti-commutation relations for $\lambda=0$. 
In Section~5, we then deform these relations away from the free field point ($\lambda=0$), using as a guiding principle our insight into the correct description of the minimal and conjugate minimal representations. We furthermore test our ansatz by comparing to the free field limit, and by showing that the additional generators lead to states in the correct representations. 
Our conclusions and avenues for future work are outlined in Section~6. 
There are two appendices: in appendix~A, we have spelled out some of the free field relations that we did not want to put in the main part of the text, and in Appendix~B we have summarized the defining relations of the supersymmtric affine Yangian we have found. 
\bigskip

\noindent {\bf Note added:} As we were in the final stages of this work we were made aware of \cite{PR} which contains some overlap with Section~2 of our paper.

\section{Building up the ${\cal N}=2$ ${\cal W}_\infty$ algebra}\label{sec:2}

In this section we explain that the  ${\cal W}^{({\cal N}=2)}_\infty[\lambda]$ algebra contains two bosonic ${\cal W}_\infty[\mu]$ algebras as mutually commuting subalgebras, one at $\mu=\lambda$ and one at $\mu=1-\lambda$. 
Note that it is known, see e.g.\ eq.~(202) in \cite{Gaberdiel:2012uj}, that the ${\cal N}=2$ higher spin algebra can be written in this manner
\be
{\rm shs}[\lambda]^{({\rm bos})} \cong {\rm hs}[\lambda] \oplus {\rm hs}[1-\lambda] \ .
\ee
However, it is not obvious whether this will also be true for the full quantum ${\cal W}_\infty^{({\cal N}=2)}[\lambda]$ algebra.\footnote{
	For the case of the ${\cal N}=2$ ${\cal W}_3$ algebra this was already noted in \cite{Romans:1991wi,Datta:2012km}; however, for general $\lambda$ this is not known.} 
This viewpoint will be important below because it will allow us to construct the full ${\cal W}^{({\cal N}=2)}_\infty[\lambda]$ algebra starting with these bosonic subalgebras.

\subsection{Decoupling the bosonic subalgebras}

In order to understand how this comes about, it is convenient to parametrise the ${\cal W}_\infty^{({\cal N}=2)}[\lambda]$ algebra as ${\cal W}^{({\cal N}=2)}_{N,k}$, i.e.\ to express both $\lambda$ and $c$ in terms of $N$ and $k$ as 
\be
c \equiv c^{({\cal N}=2)}_{N,k}  =  \frac{3Nk}{N+k+1} \ , \qquad \lambda = \frac{N}{N+k+1} \ , 
\ee
see e.g.\ \cite{Candu:2012tr} for our conventions. 
As is also explained there, the ${\cal W}^{({\cal N}=2)}_{N,k}$ algebra contains two spin $h=2$ fields: the stress energy tensor $T$, and the primary spin $h=2$ field $W\equiv W^{(2)0}$. $W$ is also primary with respect to the $\mathfrak{u}(1)$ current $J$, but $T$ is not; however, we can define the decoupled spin $2$ field via 
\be\label{Ttilde}
\tilde{T} = T - \frac{3}{2 c} \, : J J : \ . 
\ee
The modes of these two fields then satisfy the commutation relations
\begin{eqnarray}
{}[\tilde{T}_m,\tilde{T}_n] & = & (m-n) \tilde{T}_{m+n} + \frac{(c-1)}{12} \, m (m^2-1) \, \delta_{m,-n}  \nonumber \\
{}[\tilde{T}_m,W_n] & = & (m-n) W_{m+n} \\
{}[W_m,W_n] & = & (m-n) \, \left( \frac{2n_2}{(c-1)} \tilde{T}_{m+n} + \frac{c_{22,2}}{2} W_{m+n} \right)  + \frac{n_2}{6} \, m (m^2-1) \, \delta_{m,-n} \ ,  \nonumber
\end{eqnarray}
where we are using the same conventions as in \cite{Candu:2012tr}, and the last identity is directly read off from eq.~(2.14) of that paper. 
To find the two commuting Virasoro algebras, we make the ansatz 
\be\label{ansatz}
\tilde{T}_m = a^+ L^+_m + a^- L^-_m \ , \qquad
W_m = b^+ L^+_m + b^- L^-_m \ , 
\ee
and demand that $L^\pm$ commute with one another and each lead to a Virasoro algebra with central charge $c_{\pm}$ where $c_+\geq c_-$. In particular, it follows that 
\begin{equation}
(c-1) = c_+ + c_- \ , \qquad 
0 = b^+ c_+ + b^- c_- \ .
\end{equation}
These two conditions fix the coefficients $a^\pm$ and $b^\pm$ uniquely, and one finds that the two solutions are 
\be\label{abvalue}
a^\pm = 1 \ , \qquad
b^\pm = \frac{c_{22,2}}{4} \mp \frac{1}{4} \, \sqrt{(c_{22,2})^2 + 32 \frac{n_2}{(c-1)}} \ . 
\ee
The corresponding central charges are then 
\be
c_+ = \frac{(c-1) b^-}{b^- - b^+} \ , \qquad
c_- = - \frac{(c-1)b^+}{b^- - b^+} \ .
\ee
Upon plugging in the formulae for $b^\pm$  and using the explicit expression for $\gamma=(c_{22,2})^2$ from \cite{Candu:2012tr}, one finds that $c_{\pm}$ equals ${c}_{N,k}$ and ${c}_{k,N}$, respectively,  where 
\be
{c}_{N,k} =  (N-1) \Bigl[1- \frac{N(N+1)}{(N+k) (N+k+1)} \Bigr]  
\ee
is the central charge of the bosonic ${\cal W}_{N,k}$ algebra (without the additional $\mathfrak{u}(1)$ current). Note that the full (decoupled) stress energy tensor $\tilde{T}$, defined in eq.~(\ref{Ttilde}),  is indeed the sum of $L^+$ and $L^-$, 
\be
\tilde{T}_m = L^+_m + L^-_m \ ,
\ee
as also follows from (\ref{abvalue}). In particular, this implies that the total central charge of the ${\cal N}=2$ algebra must equal  --- the ``$+1$" comes from the $\mathfrak{u}(1)$ factor we have divided out --- 
\be\label{cNkN=2}
c_{N,k} + c_{k,N} + 1 = \frac{(3k-1)N - k -1 }{N+k+1} + 1 =  \frac{3Nk}{N+k+1}  = c^{({\cal N}=2)}_{N,k} \ ,
\ee
as is indeed true. 

For each integer spin $s\geq 3$, the ${\cal W}_\infty^{({\cal N}=2)}[\lambda]$ (or ${\cal W}_{N,k}^{({\cal N}=2)}$)
algebra contains at least two Virasoro primary fields. It seems plausible that among these spin $s$ fields, we can always find two fields $W^{(s)\pm}$ such that\footnote{We have checked explicitly that this is the case for spin $s=3$.} 
\be\label{assumption}
{} [ L^\pm_m,W^{(s) \pm}_n] = \bigl( (s-1)m - n) \, W^{(s) \pm}_{m+n} \ , \qquad [ L^\pm_m,W^{(s) \mp}_n] = 0 \ . 
\ee
Since all $W^{(s)+}$ fields commute with $L^-$, the same must be true for their commutator, and hence the VOA generated by $L^+$ and the $W^{(s)+}$ fields must close. (Obviously, a similar statement also holds for $L^-$ and the $W^{(s)-}$ fields.) Furthermore, the commutator of $W^{(s)+}$ with $W^{(t)-}$ must vanish since, with respect to $L^+$, say, $W^{(t)-}$ behaves like the identity field and hence does not give rise to a non-trivial commutator. 
Thus, if for each spin $s$ there are two fields such that (\ref{assumption}) holds, it follows that the ${\cal W}_\infty^{({\cal N}=2)}[\lambda]$ algebra contains two commuting bosonic ${\cal W}_{\infty}[\mu]$ algebras, one generated by the fields $L^+$ and the $W^{(s)+}$, and the other generated by $L^-$ and the $W^{(s)-}$. Given that their central charges equal $c_{N,k}$ and $c_{k,N}$, it is very plausible that the relevant ${\cal W}_{\infty}$ algebras are just the bosonic ${\cal W}_{N,k}$ and ${\cal W}_{k,N}$ algebras, respectively, i.e.\ that 
\be
{\cal W}^{({\cal N}=2)}_{N,k} \, \supset \,  {\cal W}_{N,k} \oplus {\cal W}_{k,N} \ . 
\ee
In order to confirm this we would have to construct the relevant fields and determine their $C_{33}{}^4$ structure constants, but we have not attempted to do so here. 
Note that this structure also nicely reflects the $\mathbb{Z}_2 \subset \mathbb{Z}_2\times\mathbb{Z}_2$ symmetry of the ${\cal W}^{({\cal N}=2)}_{N,k}$ algebra that is realized by  $N\leftrightarrow k$.

\subsection{Character analysis}

Next we want to understand the additional generators that need to be added to the two bosonic ${\cal W}_\infty[\mu]$ algebras in order to generate 
${\cal W}^{({\cal N}=2)}_\infty[\lambda]$. For the following it will be convenient to add a single free boson field to ${\cal W}^{({\cal N}=2)}_\infty[\lambda]$. 
Then the corresponding vaccum character equals
\be\label{chi0}
\chi_0(q) = \prod_{n=1}^{\infty} \frac{(1+q^{n+\frac{1}{2}})^{2n}}{(1-q^n)^{2n}} \ ,
\ee
since the vacuum character of the ${\cal W}_{\infty}^{({\cal N}=2)}$ algebra is 
\be\label{chi0p}
\chi_0'(q) = \prod_{s=1}^{\infty} \prod_{n=s}^{\infty} \frac{(1+q^{n+\frac{1}{2}})^2}{(1-q^n)(1-q^{n+1})} \ , 
\ee
and a single free boson contributes
\be\label{fb}
\chi_{\rm fb} = \prod_{n=1}^{\infty} \frac{1}{(1-q^n)} \ .
\ee
We want to organise this character in terms of ${\cal W}_{1+\infty}[\lambda] \oplus {\cal W}_{1+\infty}[1-\lambda]$. The va\-cuum character of each ${\cal W}_{1+\infty}[\mu]$ algebra is described by a plane partition, see e.g.\ \cite{Prochazka:2015deb,Gaberdiel:2017dbk}
\be
\chi_{\rm pp} = \prod_{n=1}^{\infty} \frac{1}{(1-q^n)^n} \ , 
\ee
and hence the two algebras account precisely for the denominator of $\chi_0(q)$ in eq. (\ref{chi0}). The numerator of (\ref{chi0})  corresponds to the fermionic excitations, and is accounted for in terms 
of bi-minimal representations of the two algebras\footnote{The bi-minimal representations that are relevant here are ``minimal" with respect to one factor, and ``conjugate-minimal" with respect to the other. There are therefore two such representations, namely ``minimal"--``conjugate-minimal" and ``conjugate-minimal"--``minimal".} together with their tensor powers. Since for each representation $R$ of ${\cal W}_{1+\infty}[\mu]$ the character is of the form
\be
\chi_R^{({\rm wedge})}(q) \, \chi_{\rm pp}(q) \ , 
\ee
this amounts to the condition that 
\be\label{charid}
\prod_{n=1}^{\infty} (1+q^{n+\frac{1}{2}})^{2n} = \sum_{R}  \chi_R^{({\rm wedge})\,  [\lambda]}(q) \cdot \chi_{\bar{R}^T}^{({\rm wedge})\, [1-\lambda]}(q) \ ,
\ee
where $R$ runs over all representations that appear in finite tensor powers of the two bi-minimal representations, and $\bar{R}^T$  is the conjugate representation to $R^T$, with $T$ denoting the transpose of $R$. Since $R$ involves in general box and anti-box representations, and since the wedge character of such a mixed representation is simply the product of the wedge character of the box representation and that of the anti-box representation, the above identity follows from 
\be\label{claim}
\prod_{n=1}^{\infty} (1+y q^{n+\frac{1}{2}})^{n} = \sum_{S} y^{|S|}\, \chi_S^{({\rm wedge})\,  [\lambda]}(q) \cdot \chi_{S^T}^{({\rm wedge})\, [1-\lambda]}(q) \ ,
\ee
where $S$ runs over all Young diagrams (labelling say box-representations), $S^T$ is the transpose Young diagram (labelling now anti-box representations), and $|S|$ denotes the number of boxes in $S$. The first few cases are explicitly  (see \cite{Gaberdiel:2015wpo} for the general method for how to derive them)
\begin{eqnarray}
\chi^{({\rm wedge})}_{{\tiny \yng(1)}} (q) & = & \frac{q^h}{1-q} \\
\chi^{({\rm wedge})}_{{\tiny \yng(2)}} (q) & = & \frac{q^{2h}}{(1-q)(1-q^2)} \\
\chi^{({\rm wedge})}_{{\tiny \yng(1,1)}} (q) & = & \frac{q^{2h+1}}{(1-q)(1-q^2)} \\
\chi^{({\rm wedge})}_{{\tiny \yng(3)}} (q) & = & \frac{q^{3h}}{(1-q)(1-q^2)(1-q^3)} \\
\chi^{({\rm wedge})}_{{\tiny \yng(2,1)}} (q) & = & \frac{q^{3h+1}}{(1-q)^2(1-q^3)} \\
\chi^{({\rm wedge})}_{{\tiny \yng(1,1,1)}} (q) & = & \frac{q^{3h+3}}{(1-q)(1-q^2)(1-q^3)}  \ .
\end{eqnarray}
It is then straightforward to check (\ref{claim}) explicitly,\footnote{We have done this up to $q^{10}$. It should not be too hard to prove this analytically, but we have not attempted to do so.}
provided that we take the box representations of ${\cal W}_{1+\infty}[\lambda]$ and ${\cal W}_{1+\infty}[1-\lambda]$ to be the ones with conformal dimensions
\be\label{confd}
h = \frac{1}{2} (1+\lambda) \ , \qquad \hat{h} = \frac{1}{2} \bigl( 1 + (1-\lambda) \bigr) \ , 
\ee
respectively, so that the total conformal dimension equals
\be\label{confsuper}
h + \hat{h} = \frac{1}{2} (1+\lambda) + \frac{1}{2} \bigl( 1 + (1-\lambda) \bigr) = \frac{3}{2} \ , 
\ee
thus reproducing the conformal dimension of the supercharge.

\section{The minimal and conjugate minimal representation}

As we have seen in the previous section, the additional generators that need to be added to the two bosonic ${\cal W}_{1+\infty}$ algebras transform in bi-minimal representations with respect to these two algebras. For the following it will be important to describe these representations from the viewpoint of the affine Yangian.

\begin{figure}[h!]
	\centering
	\includegraphics[trim=0cm 20cm 3cm 4cm, width=1.0\textwidth]{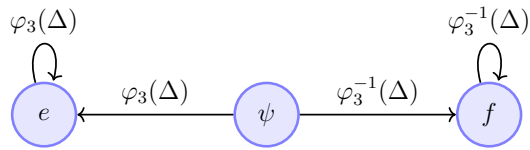}
	\caption{The OPE relations of the affine Yangian.	}
	\label{figOPEbosonic}
\end{figure}

Recall from \cite{Prochazka:2015deb,Gaberdiel:2017dbk} that the defining relations of the affine Yangian can be written as 
\be\label{efrel}
e(z)\, f(w) - f(w)\, e(z) = - \frac{1}{\sigma_3}\, \frac{\psi(z) - \psi(w)}{z-w} \ , 
\ee
and
\begin{align}
e(z)\, e(w) \, \ \sim \ \ & \varphi_3(z-w)\, e(w)\, e(z)  \label{eegen} \\
f(z)\, f(w) \, \ \sim \ \ & \varphi_3^{-1}(z-w)\, f(w)\, f(z) \label{ffgen} \\
\psi(z)\, e(w) \, \ \sim \ \ & \varphi_3(z-w)\, e(w)\, \psi(z) \label{psiegen} \\
\psi(z)\, f(w) \, \ \sim \ \ & \varphi_3^{-1}(z-w)\, f(w)\, \psi(z) \ , \label{psifgen} 
\end{align}
where `$\sim$' means equality up to terms that are regular at $z=0$ or $w=0$. Here the fields are expanded in terms of modes as 
\be\label{generating}
e(z) = \sum_{j=0}^{\infty} \, \frac{e_j}{z^{j+1}} \ , \qquad 
f(z) = \sum_{j=0}^{\infty} \, \frac{f_j}{z^{j+1}} \ , \qquad 
\psi(z)  = 1 + \sigma_3 \, \sum_{j=0}^{\infty} \frac{\psi_j}{z^{j+1}} \ , 
\ee
and  the function $\varphi_3(z)$ is defined by 
\be\label{varphidef}
\varphi_3(z) = \frac{(z+h_1) (z+h_2) (z+h_3)}{(z-h_1) (z-h_2) (z-h_3)}  = \frac{z^3 + \sigma_2 z + \sigma_3}{z^3 + \sigma_2 z - \sigma_3}\ .
\ee
The $h_i$ parameters satisfy $h_1+h_2+h_3=0$, and we have defined 
\be\label{sigma23}
\sigma_2 = h_1 h_2 + h_2 h_3 + h_1 h_3 \ , \qquad \sigma_3 = h_1 h_2 h_3 \ . 
\ee
The structure of these OPEs can be summarised by the diagram of Fig.~\ref{figOPEbosonic}. 

In terms of the conformal field theory language, the $h_i$ parameters and $\psi_0$ can be expressed as, 
\be
\psi_0 = N 
\ee
and 
\be
h_1 =  -\sqrt{\frac{N+k+1}{N+k}} \ , \qquad h_2 =  \sqrt{\frac{N+k}{N+k+1}} \ , \qquad h_3 = \frac{1}{\sqrt{(N+k)(N+k+1)}} \ , \label{hkn}
\ee
see eqs.~(3.51)  and (3.52) of \cite{Gaberdiel:2017dbk}.

As is explained in \cite{Prochazka:2015deb,Gaberdiel:2017dbk}, the different states of the vacuum representation are described by plane partitions where the eigenvalues of $\psi_i$ on the configuration $\Lambda$ are given by 
\be\label{psieig}
\psi_\Lambda(z) = \bigl( 1 + \frac{\psi_0 \sigma_3}{z} \bigr) \, \prod_{ {\tiny \yng(1)} \in (\Lambda)} \varphi_3(z - h({\tiny \yng(1)}) ) \ , 
\ee
where 
\be\label{hbox}
h({\tiny \yng(1)}) = h_1 x({\tiny \yng(1)}) + h_2 y({\tiny \yng(1)}) +  h_3 z({\tiny \yng(1)}) 
\ee
with $x({\tiny \yng(1)})$ the $x$-coordinate of the box, etc. Furthermore, the representations of the affine Yangian are parametrised by non-trivial asymptotic box stackings, and the charges of the corresponding states are still given by (\ref{psieig}), except that now the infinite product (over the infinitely many boxes defining the asymptotic configuration) must be suitably regularized. 

\subsection{The minimal representation}

The simplest non-trivial representations are the minimal representations whose asym\-ptotic box configuration consists of a single row of boxes extending along either $x_1$, $x_2$ or $x_3$. For our analysis above, the minimal representation corresponding to an asymptotic single box in the $x_2$ direction will play a central role.\footnote{It is the one whose conformal dimension equals (\ref{confd}) in the classical limit. Note that selecting out $x_2$ breaks the $S_3$ symmetry of the affine Yangian to $\mathbb{Z}_2$.} Its ground state has the charges 
\begin{align}
\psi(u)  & =   \bigl( 1 + \frac{\psi_0 \sigma_3}{u} \bigr) \, \prod_{n=0}^{\infty} \varphi_3(u - n h_2) \nonumber \\ 
& =  \bigl( 1 + \frac{\psi_0 \sigma_3}{u} \bigr) \, \frac{u (u+h_2)} {(u-h_1) (u-h_3)} \ . \label{psiudef}
\end{align}
Expanding out in inverse powers of $u$, this is of the form
\be\label{psiu}
\psi(u) = 1 + \frac{\psi_0 \sigma_3}{u} - \frac{h_1 h_3}{u^2} + \frac{h_1 h_3 (h_2 - \psi_0 \sigma_3)}{u^3} + {\cal O}(u^{-4} ) \ . 
\ee
Comparing with (\ref{generating}) we read off 
\be\label{charges}
\psi_1 =-\frac{1}{h_2} \ , \qquad \psi_2 = \frac{h_2-\s_3 \psi _0}{h_2 } \ , \qquad 
\psi_3  = \frac{ h_2 \s_3   \psi _0 -h_1^2-h_3 h_1-h_3^2}{h_2 } \ . 
\ee

\subsection{Charges of the conjugate representation}

As will become clear from the free field analysis of the next section, the above minimal representations will not suffice. In fact, the 
bi-minimal representations that appear in (\ref{charid}) have the property that they are minimal with respect to one ${\cal W}_\infty$ algebra, but conjugate minimal (or anti-minimal) with respect to the other. Thus in order to describe these generators we also need to understand how to describe the conjugate minimal representation from the affine Yangian perspective. In the following we shall make a general proposal for how this works. 

Given a box representation described by $\varphi(u)$ (where again $\varphi(u)$ does not include the vacuum factor $\psi_0(u)$), we claim that the corresponding conjugate representation has charges given by 
\be\label{conjchg}
\varphi^{-1}(-u - \psi_0\sigma_3) \ . 
\ee
The signficance of this shift is that it turns the vacuum factor $\psi_0(u) = (1 + \frac{\psi_0\sigma_3}{u})$ into 
\be
\psi_0(-u- \psi_0\sigma_3) = \psi_0^{-1}(u) \ .
\ee
Thus the full eigenvalue function of the anti-box representation is 
\be
\psi_0(u) \, \varphi^{-1}(-u - \psi_0\sigma_3) = \bigl( \psi_0(v) \, \varphi(v) \bigr)^{-1} \ , \qquad 
\hbox{where $v= - u - \psi_0 \sigma_3\ $,}
\ee
and therefore indeed just the inverse of $\psi_\Lambda(u) = \psi_0(u) \varphi(u)$. (Note that the shift (and sign) transformation from $u$ to $v$ is just the spectral flow 
(and scaling) automorphism of the affine Yangian, see e.g.\ sections~2.2 and 2.3 of \cite{Prochazka:2015deb}.)

We can check this proposal explicitly by checking whether the first few  $W^s_0$ charges change correctly --- recall that for the conjugate representation the charges of the odd spin $W^s_0$ generators must have the opposite sign, while the even spin generators are the same. Suppose then that $\varphi(u)$ describes a given representation with eigenvalues $\psi_j$, i.e.\ we have the expansion
\be
(1 + \tfrac{\psi_0\sigma_3}{u})\,  \varphi(u) = 1 + \frac{\psi_0\sigma_3}{u} + \frac{\psi_1\sigma_3}{u^2} + \frac{\psi_2\sigma_3}{u^3} + \frac{\psi_3\sigma_3}{u^4} + \cdots \ .
\ee
Then, according to the above proposal, the power series expansion of the conjugate representation is 
\be
(1 + \tfrac{\psi_0\sigma_3}{u})\,  \varphi^{-1}(-u - \psi_0\sigma_3) = 1 + \frac{\psi_0\sigma_3}{u} - \frac{\psi_1\sigma_3}{u^2}  +  \frac{\psi_2\sigma_3}{u^3} 
+ \frac{\sigma_3  ( - \psi_3 - \psi_0 \psi_2 \sigma_3 +\psi_1^2 \sigma_3)}{u^4} + \cdots
\ee
This predicts that the conjugate representation has the charges
\be
\bar{\psi}_0 = \psi_0\ , \qquad 
\bar{\psi}_1 = - \psi_1 \ , \qquad \bar{\psi}_2 = \psi_2 \ , \qquad \bar{\psi}_3 = - \psi_3 - \psi_0 \psi_2 \sigma_3 +\psi_1^2 \sigma_3 \ . 
\ee
Together with the form of the spin $s=3$ charge from \cite{Gaberdiel:2017dbk}
\bal
W^3_0&=-\frac{1}{3}\psi_3-\frac{\s_3\psi_0}{6}\psi_2+\frac{\s_3}{6}\psi_1\psi_1\ ,
\eal
it follows that the value of $W^3_0$ on the conjugate representation equals 
\begin{eqnarray}
\bar{w} & = &  -\frac{1}{3}\bar{\psi}_3-\frac{\s_3 \bar{\psi}_0}{6}\bar{\psi}_2+\frac{\s_3}{6}\bar{\psi}_1\bar{\psi}_1 \nonumber \\
& = & - \frac{1}{3} \bigl( - \psi_3 - \psi_0 \psi_2 \sigma_3 +\psi_1^2 \sigma_3 \bigr) -\frac{\s_3 {\psi}_0}{6} {\psi}_2+\frac{\s_3}{6}{\psi}_1{\psi}_1 \nonumber \\ 
& = &  - \Bigl[ - \frac{1}{3}  \psi_3-\frac{\s_3\psi_0}{6}\psi_2+\frac{\s_3}{6}\psi_1\psi_1 \Bigr] = - w \ ,
\end{eqnarray}
as expected. This is a fairly non-trivial consistency check of this proposal. 
\smallskip

\noindent In particular, for the conjugate minimal representation from above we find 
\be\label{anticharges1}
\bar{\psi}_1= \frac{1}{h_2 }\ , \qquad 
\bar{\psi}_2=1- h_1 h_3\psi _0 \ , \qquad 
\bar{\psi}_3={h_2 \left(1-h_1 h_3 \psi _0 \right){}^2} \ ,
\ee
while for $s\geq 3$ we find the simple closed form expression
\be
\bar{\psi}_s={h_2^{s-2} \left(1-h_1 h_3 \psi _0 \right){}^{s-1}}\ .
\ee

\subsection{The conformal dimensions at finite $N$ and $k$}\label{sec:finiteconf}

As an aside, the above analysis now allows us to check the finite $N$ and $k$ corrections to the conformal dimension (\ref{confsuper}). 
According to \cite{Prochazka:2015deb,Gaberdiel:2017dbk}, the conformal dimensions with respect to the coupled theory (where the $\mathfrak{u}(1)$ generator has not been removed) equals, see eq.~(\ref{charges}) 
\be\label{heval1}
h = \frac{1}{2} \psi_2 = \frac{1}{2} \bigl(1 - \psi_0 h_1 h_3\bigr)  = \frac{1}{2} \Bigl(1 + \frac{N}{N+k}\Bigr)  \ , 
\ee
where we have used the dictionary (\ref{hkn}). Note that this is true both for the minimal representation, as well as the anti-minimal representation, see eq.~(\ref{anticharges1}). We should also mention in passing that the decoupled conformal dimension is then, see eq.~(5.67) of \cite{Prochazka:2015deb}
\be
h_{\rm dec} =  \frac{1}{2} \, \psi_2 - \frac{\psi_1^2}{2 \psi_0} = \frac{(N-1)}{2N} \, \Bigl( 1 + \frac{N+1}{N+k} \Bigr) \ , 
\ee
and hence agrees indeed with the conformal dimension of $h({\rm f};0)$ in the coset, see, e.g., eq.~(2.13) of \cite{Gaberdiel:2010pz}. 

For the problem at hand, however, we should work with the coupled conformal dimension (since we are dealing with ${\cal W}_{1+\infty}$ rather than just ${\cal W}_{\infty}$). Furthermore, for the ${\cal N}=2$ construction, we consider two plane partitions that correspond to $\lambda$ and $1-\lambda$, i.e., for the second ${\cal W}_\infty$ algebra we should exchange the roles of $N$ and $k$. Note that, according to the dictionary of eq.~(\ref{hkn}), this only affects $\psi_0$, but not the values of $h_i$. Thus we shall work with the same values of $h_i$ for both affine Yangians, but distinguish the two affine Yangians by setting 
\be\label{psi0hatpsi0}
\psi_0 = N \ , \qquad \hat{\psi}_0 = k \ . 
\ee
Note that, using (\ref{hkn}), this relation can be written as 
\be\label{hatpsi0}
h_1 h_3 \psi_0 = - \frac{N}{N+k} = - 1 + \frac{k}{N+k} = - 1 - h_1 h_3 \hat{\psi}_0  \ , 
\ee
or equivalently as 
\be\label{magic}
\sigma_3 \hat{\psi}_0 = - h_2 - \sigma_3 \psi_0 \ . 
\ee

If we then add (\ref{heval1}) to the corresponding expression with $N \leftrightarrow k$, we find altogether
\be\label{N2conf}
h_{\rm tot}  =  \frac{1}{2} \bigl(1 - \psi_0 h_1 h_3\bigr)  + \frac{1}{2} \bigl(1 - \hat{\psi}_0 h_1 h_3\bigr)  = \frac{3}{2} \ , 
\ee
as desired. This is another highly non-trivial consistency check for this construction to work also at the quantum level. 

It may also be worth noting that the $\mathbb{Z}_2\times \mathbb{Z}_2$ symmetry of the ${\cal W}^{({\cal N}=2)}_\infty$ algebra \cite{Candu:2012tr} has a nice geometric interpretation in this setting. First of all, the $S_3$ symmetry of each of the two affine Yangians is broken down to a common $\mathbb{Z}_2$ symmetry that exchanges the $x_1$ and $x_3$ direction. In addition, there is the symmetry exchanging the roles of the two affine Yangians, which corresponds to the $N \leftrightarrow k$ transformation.

\subsubsection{Representations}

One can similarly understand how the two minimal ${\cal N}=2$ representations appear from the above bosonic viewpoint. The relevant representations have conformal dimensions 
\be\label{N2reps}
h({\rm f};0,N) = \frac{N}{2(N+k+1)} \ , \qquad \hbox{and} \qquad h\bigl(0;{\rm f},-(N+1)\bigr) = \frac{k}{2(N+k+1)} \ , 
\ee
see, e.g.\ eq.~(3.9) of \cite{Candu:2012tr}. From the above perspective, these representations correspond to the representation that has an infinite row of boxes along the $x_1$ direction for either of the two plane partitions. Indeed, it follows from \cite{Prochazka:2015deb,Gaberdiel:2017dbk} that the relevant conformal dimensions are (cf.\ eq.~(\ref{heval1}))
\be\label{heval2}
h^{(1)}_{\rm min} = \frac{1}{2} \bigl(1 - \psi_0 h_2 h_3\bigr)  = \frac{k+1}{2 (N+k+1)} 
\ee
and
\be\label{heval2p}
{h}^{(2)}_{\rm min}  = \frac{1}{2} \bigl(1 - \hat{\psi}_0 {h}_2 {h}_3\bigr)  = \frac{N+1}{2 (N+k+1)}  \ .
\ee
Note that these conformal dimensions are higher than those in (\ref{N2reps}), with the difference in both cases being equal to 
\be
\delta h = \frac{1}{2(N+k+1)} \ . 
\ee
This is the contribution of the overall $\mathfrak{u}(1)$ generator that was added in (\ref{fb}). In general, the representations of the ${\cal N}=2$ affine Yangian will therefore be described by infinite box configurations extending in the $x_1$ and $x_3$ direction for both plane partitions. (The representations where the boxes extend along the $x_3$ direction are again the non-perturbative representations, in analogy to what happens in the bosonic case, see \cite{Gaberdiel:2012ku}.)

\section{The Yangian at the free field point}

For the following it will sometimes be important to compare our ansatz with an explicit free field realization. Recall that at 
$\lambda=0$ (or $\lambda=1$), the ${\cal W}_\infty^{({\cal N}=2)}$ algebra has a free field construction in terms of free complex fermions and bosons. More explicitly, the neutral bilinears in the fermions, i.e.\ the fields of the form 
\be
\sum_j \partial^l \chi^j  \, \partial^m \bar{\chi}^j  
\ee
generate ${\cal W}_{1+\infty}[0]$, while the neutral bilinears in the bosons, i.e.\ the fields of the form 
\be
\sum_j \partial^l \phi^j  \, \partial^m \bar{\phi}^j  
\ee
give rise to ${\cal W}_\infty[1]$. (Here $\chi^j$ and $\bar{\chi}^j$ are the complex fermions, and $\phi^j$ and $\bar\phi^j$  are the complex bosons.)  On the other hand, the fermionic generators are linear combinations of the form
\be
\sum_j \partial^l \chi^j  \, \partial^m \bar{\phi}^j  \qquad \hbox{and} \qquad  \sum_j \partial^l \bar\chi^j \, \partial^m \phi^j \ . 
\ee
From the viewpoint of the two bosonic ${\cal W}_\infty$ algebras, i.e.\ ${\cal W}_{1+\infty}[0]$ and ${\cal W}_\infty[1]$, these generators transform in the `bi-minimal' representation. Indeed, the fermion field $\chi^j$ corresponds (with respect to ${\cal W}_{1+\infty}[0]$) to the representation $({\rm f};0)$ in the coset language, while the boson field $\partial\phi^j$ describes (with respect to ${\cal W}_{\infty}[1]$) the representation $({\rm f};0)$. Thus the two fermionic fields above correspond to the states in 
\be\label{fermions}
({\rm f};0) \otimes (\bar{{\rm f}};0) \qquad \hbox{and} \qquad (\bar{\rm f};0) \otimes ({\rm f};0)  \ ,
\ee
respectively. (Here we have used that the complex conjugate fermion transforms in $(\bar{\rm f};0)$, and correspondingly for the complex conjugate boson.)

In terms of the description in terms of plane partitions, this means that the first fermionic generators act as an addition of an infinite row of boxes with respect to the first plane partition --- the one corresponding to ${\cal W}_{1+\infty}[0]$ --- while it acts as an an addition of an infinite row of anti-boxes with respect to the second plane partition --- the one corresponding to ${\cal W}_\infty[1]$. 

\subsection{The affine Yangian generators at $\lambda=0$}

We can use this free field realisation to construct the relevant affine Yangian generators for this special case, and work out their commutation and anti-commutation relations. In the next section we shall explain how to modify these relations as we move away from $\lambda=0$. 

We recall  from \cite{Gaberdiel:2017dbk} (see also \cite{Prochazka:2015deb})  that for the bosonic ${\cal W}_{1+\infty}[0]$ algebra, the corresponding affine Yangian generators can be defined as 
\begin{eqnarray}\label{psidef}
\psi_r & = & {\displaystyle  \sum_{m\in\mathbb{Z}+\frac{1}{2}} \sum_{i} \Bigl( (-m-\tfrac{1}{2})^r - (-m+\tfrac{1}{2})^r \Bigr)\, :\bar{\chi}^i_{-m}\chi^i_m :}   \\ 
e_r & = & -{\displaystyle \sum_{m\in\mathbb{Z}+\frac{1}{2}} \sum_{i}  \bigl(-m-\tfrac{1}{2})^r\, :\bar{\chi}^i_{-m-1} \chi^i_m : } \\
f_r & = & {\displaystyle  \sum_{m\in\mathbb{Z}+\frac{1}{2}} \sum_{i} \bigl(-m+\tfrac{1}{2})^r\, :\bar{\chi}^i_{-m+1} \chi^i_m : \ , } 
\end{eqnarray}
where the free fermion modes are denoted by $\chi^i_m$ and $\bar{\chi}^i_m$. 
As was shown there, these generators satisfy the affine Yangian algebra of \cite{Prochazka:2015deb} for $\sigma_2=-1$ and $\sigma_3=0$ with $\sigma_3 \psi_0=0$. In terms of the $h_i$ parameters, this corresponds to the case
\be\label{hdef}
h_1 = - 1 \ , \qquad h_2 = 1 \ , \qquad h_3 =0 \ ,
\ee
see eq.~(\ref{hkn}) above. We also need a description for the affine Yangian generators associated to ${\cal W}_{\infty}[1]$, and they are given as 
\begin{eqnarray}\label{hated}
\hat{\psi}_r & = &  {\displaystyle   \sum_{m\in\mathbb{Z}}  \sum_{i} \Bigl( (  m + 1) (-m)^{r-2} + \bigl(-m+1 \bigr)^{r-1} \Bigr)\, 	:   \bar\alpha^i_{-m} \alpha^i_{m}:  } \\
\hat{e}_r & = & {\displaystyle   \sum_{m\in\mathbb{Z}}  \sum_{i} (-m)^{r-1}\, :    \bar\alpha^i_{-1-m} \alpha^i_m :  } \\
\hat{f}_r & = & {\displaystyle -  \sum_{m\in\mathbb{Z}} \sum_{i} \bigl(-m+1\bigr)^{r-1}\, :  \bar\alpha^i_{1-m}  \alpha^i_{m} : \ . } \label{hated3}
\end{eqnarray}
This leads to the affine Yangian with $h_i$ being given by (\ref{hdef}); the only difference to the case of ${\cal W}_\infty[0]$ above is that now $\sigma_3 \hat\psi_0=-1$. On the face of it, these definitions only make sense for $\hat{e}_r$, $\hat{f}_r$ with $r\geq 1$ and $\hat{\psi}_r$ with $r\geq 2$. However, we can at least formally extend these definitions to include also $\hat{e}_0$ and $\hat{\psi}_1$ by setting $\alpha^i_0\equiv 0$, i.e., by dropping the term with $m=0$ from all of these expressions. (Similarly, we could define $\hat{f}_0$ by setting $\bar{\alpha}^i_0\equiv 0$.)
The generator $\hat{e}_0$ is then the $-1$ mode of a non-local field with spin $1$. One checks by an explicit calculation that it satisfies the correct commutation relation with the $\hat{\psi}_r$ modes, in particular (see eq.~(4.13) of \cite{Gaberdiel:2017dbk})
\be
{}[\hat{\psi}_1,\hat{e}_0] = 0 \ , \qquad 
{}[\hat{\psi}_2,\hat{e}_0] =  2 \hat{e}_0 \ , \qquad 
{}[\hat{\psi}_3,\hat{e}_0] =  6 \hat{e}_1 - 2 \hat{e}_0 \ . 
\ee
For the fermionic generators we now make the ansatz 
\be\label{xdef}
x_s = \sum_{m\in\mathbb{Z} } \sum_i (-m-1)^{s-\frac{1}{2}} \, \, :\bar\chi^i_{-\frac{3}{2}-m} \, {\alpha}^i_{m}: \  ,
\ee
and 
\be\label{barx}
\bar{x}_s =  \sum_{m\in\mathbb{Z}} \sum_i (m+1)^{s-\frac{1}{2}} \, \, :{\chi}^i_{-\frac{3}{2}-m} \, \bar{\alpha}^i_{m}: \  ,
\ee
where $s=\frac{1}{2}, \frac{3}{2}, \ldots$, and we define the generating functions by 
\be
x(z) = \sum_{s=1/2}^{\infty} x_s\,  z^{-s-1/2} \ , \qquad
\bar{x}(z) = \sum_{s=1/2}^{\infty} \bar{x}_s \, z^{-s-1/2} \ .
\ee
Note that 
\begin{equation}
x_s \sim W^{(s+1)+}_{-3/2} \ , \qquad \hbox{and} \qquad \bar{x}_s \sim W^{(s+1)-}_{-3/2}\ , 
\end{equation}
 i.e.\ the term with $s=\frac{1}{2}$ corresponds to the supercharge, etc. 
Obviously, the algebra also contains the corresponding $+3/2$ modes, which we may define via conjugation as  
\be
{y}_s = \sum_{m\in\mathbb{Z} } \sum_i (m-1)^{s-\frac{1}{2}} \, \, :{\chi}^i_{\frac{3}{2}-m} \, \bar{\alpha}^i_{m}: \ , 
\ee
and
\be\label{ydef}
\bar{y}_s = \sum_{m\in\mathbb{Z} } \sum_i (1-m)^{s-\frac{1}{2}} \, \, :\bar{\chi}^i_{\frac{3}{2}-m} \, {\alpha}^i_{m}: \  ,
\ee
where the corresponding generating functions are 
\be
y(z) = \sum_{s=1/2}^{\infty} y_s\,  z^{-s-1/2} \ , \qquad
\bar{y}(z) = \sum_{s=1/2}^{\infty} \bar{y}_s \, z^{-s-1/2} \ ,
\ee
and
\be
x_s^\dagger = {y}_s \ , \qquad \hbox{and} \qquad \bar{x}_s^\dagger = \bar{y}_s \ . 
\ee
Their treatment is similar to that of the $x_r$ and $\bar{x}_r$ generators, and is therefore relegated to Appendix~\ref{app:Serre}.
\smallskip

It is now straightforward to work out the commutation and anti-commutation relations of these generators. For example, one finds 
\bal
0 & = [\psi_{r+2},{x}_s]-2[\psi_{r+1},{x}_{s+1}] +[\psi_r,{x}_{s+2 }]+[\psi_{r+1},{x}_{s }]-[\psi_r,{x}_{s+1 }] \label{psix}\\
0 & = [e_{r+1},x_s]-[e_r,x_{s+1}] + [e_r,x_{s }]  \label{exmain}\\
0 & = [{f}_{r+1 },x_s] -  [{f}_r,x_{s+1 }]   \ , \label{fxmain}
\eal
as well as 
\bal
0 & = [\psi_{r+2},\bar{x}_s]-2[\psi_{r+1},\bar{x}_{s+1}] +[\psi_r,\bar{x}_{s+2 }]-[\psi_{r+1},\bar{x}_{s }]+[\psi_r,\bar{x}_{s+1 }]  \label{psibarx}\\
0 & = [{e}_{r+1},\bar x_s]-[{e}_r,\bar x_{s+1}] - [{e}_r,\bar x_{s }] \label{ebarx} \\
0 & = [{f}_{r+1 },\bar  x_s] -  [{f}_r,\bar x_{s +1}] \ . \label{fbarx}
\eal
On the other hand, the commutation relations with the hatted modes are 
\bal
0 & = [\hat\psi_{r+2},{x}_s]-2[\hat\psi_{r+1},{x}_{s+1}] +[\hat\psi_r,{x}_{s+2 }]-3 [\hat\psi_{r+1},{x}_{s }]+3 [\hat\psi_r,{x}_{s+1 }] +2 [\hat{\psi}_r,x_s]  \label{hatpsix}\\
0 & = [\hat{e}_{r+1},x_s]-[\hat{e}_r,x_{s+1}] - 2 [\hat{e}_r,x_{s }]  \label{hatexmain}\\
0 & = [\hat{f}_{r+1 },x_s] - [\hat{f}_r,x_{s+1 }] - [\hat{f}_r,x_s]   \label{hatfxm} 
\eal
and 
\bal
0 & = [\hat\psi_{r+2},\bar{x}_s]-2[\hat\psi_{r+1},\bar{x}_{s+1}] +[\hat\psi_r,\bar{x}_{s+2 }]+[\hat\psi_{r+1},\bar{x}_{s }]-[\hat\psi_r,\bar{x}_{s+1 }]  \label{hatpsibarx} \\
0 & = [\hat{e}_{r+1},\bar x_s]-[\hat{e}_r,\bar x_{s+1}] + [\hat{e}_r,\bar x_{s }] =  0 \\
0 & = [\hat{f}_{r+1 },\bar  x_s] -  [\hat{f}_r,\bar x_{s +1}]  = 0 \ .
\eal
In addition, there are the initial conditions 
\be\label{ini}
\begin{array}{rclrclrcl}
{} [\psi_0,x_s] & = & 0 \qquad \qquad & [\psi_{1},{x}_s]&=& - {x}_s \qquad \qquad & & & \\ 
{}  [\psi_0,\bar{x}_s] & = & 0 \qquad \qquad & [\psi_{1},\bar{x}_s]&=&  \bar{x}_s \qquad \qquad & & &   \\ {}[\hat\psi_{1},{x}_s]&=&  x_s  \qquad \qquad & [\hat\psi_{2},{x}_s]&=& 2 {x}_{s} + 2 {x}_{s+1}\qquad \qquad & [\hat\psi_{3},{x}_s]&=&  
4 x_{s} + 7 {x}_{s+1} + 3 {x}_{s+2}\\
{}[\hat\psi_{1},\bar{x}_s]&=&  - \bar{x}_s \qquad\qquad & [\hat\psi_{2},\bar{x}_s]&=& 2 \bar{x}_s -2\bar{x}_{s+1}  
\qquad\qquad & [\hat\psi_{3},\bar{x}_s]&=& -2 \bar{x}_s + 5 \bar{x}_{s+1}-3\bar{x}_{s+2}\ .
\end{array}
\ee
Note that the hatted generators only start with $\hat{\psi}_2$, i.e.\ the modes $\hat{\psi}_0$ and $\hat{\psi}_1$ are initially not defined in (\ref{hated}). We have added the mode $\hat{\psi}_1$ by hand --- the result also agrees with what one obtains upon extending the definition of $\hat{\psi}_r$ in (\ref{hated}) to $r=1$, see the comments above --- and defined it so that it satisfies (\ref{hatpsix}) and (\ref{hatpsibarx}) for $r\geq 1$. 
However, (\ref{hatpsix}) and (\ref{hatpsibarx}) are then not compatible with $[\hat{\psi}_0,x_s]=0$ and $[\hat{\psi}_0,\bar{x}_s]=0$. The reason for this will become clear below: the deformed relations, see appendix~\ref{app:B2}, contain an additional contribution that survives (for $r=0$) since $\sigma_3\hat{\psi}_0=-1$. With this correction term the above results are then also compatible with the recursion relations of appendix~\ref{app:B2} for $r=0$. 
For the unhatted modes, these problems do not arise, and in fact the commutator with $\psi_2$ is determined using (\ref{psix}) and (\ref{psibarx}) with $r=0$ as 
\be
[\psi_{2},{x}_s]= {x}_s-2{x}_{s+1} \ , \qquad [\psi_{2},\bar{x}_s]=  \bar{x}_s+2\bar{x}_{s+1} \ . 
\ee
\smallskip

\noindent  Finally, for the anti-commutator of the $x_s$ and $\bar{x}_r$ 
we find 
\be\label{xbarx}
0=\{x_{i+2},\bar{x}_{j}\}-2\{x_{i+1},\bar{x}_{j+1}\}+\{x_{i},\bar{x}_{j+2}\} +  \{x_{i+1},\bar{x}_{j}\} - \{x_{i},\bar{x}_{j+1}\} - 2 \{ x_i,\bar{x}_j\} \ . 
\ee
It is also convenient to define 
\be\label{Pdef}
\{x_r,y_s\} = P_{r+s} \ ,
\ee
with the initial condition that 
\be\label{Pini}
P_1 = \frac{1}{2} \bigl( \psi_2 + \hat{\psi}_2) + \frac{3}{2} \psi_1 + N \ ,
\ee
where $N$ is the number of complex free bosons and fermions.
The $P_r$ modes satisfy a number of relations that are also spelled out in Appendix~\ref{app:Serre}.

\subsection{Identifying the representations}

The discussion around eq.~(\ref{fermions}) suggests that the fermionic generators transform in a minimal representation with respect to one ${\cal W}_\infty$ algebra, but in the conjugate minimal with respect to the other. We can now verify this also more explicitly. 

Let us first analyse the generators described by $x$. The eigenvalues of $\psi_r$ on the state $\bar{\chi}^i_{-1/2} |0\rangle$ --- this is the relevant state for the description of $x_{1/2} |0\rangle$ --- equals 
\be
\psi_1 = - 1 \ , \qquad \psi_2 = 1 \ , \qquad \psi_3 =  - 1 \ ,
\ee
where the first few $\psi_r$ generators are explicitly, see eq.~(\ref{psidef}) 
\begin{eqnarray}
\psi_1 & = &  - \sum_{m\in\mathbb{Z}+ \frac{1}{2}} \sum_i \,  :\bar\chi^i_{-m}\chi^i_{m} :  \\
\psi_2 & = & \sum_{m\in\mathbb{Z}+ \frac{1}{2}} \sum_i \, 2m\, :\bar\chi^i_{-m}\chi^i_{m} : \\
\psi_3 & = &  - \sum_{m\in\mathbb{Z}+ \frac{1}{2}} \sum_i \, \bigl( 3m^2 + \tfrac{1}{4} \bigr)\,  :\bar\chi^i_{-m}\chi^i_{m} : 
\end{eqnarray}
Using that $h_1= - 1$, $h_2=1$ and $h_3=0$ (with $\s_3\psi_0=0$) this agrees then with the charges of the minimal representation, see (\ref{charges}).

On the other hand, the charges of the state $\chi^i_{-1/2}|0\rangle$ (that is relevant for the description of $\bar{x}_s$) are 
\be\label{unhattedbarxcharges}
\psi_1(\bar{x}) = 1 \ , \qquad {\psi}_2(\bar{x}) =  1 \ , \qquad {\psi}_3(\bar{x}) =  1 \ . 
\ee
These are not the charges of the minimal representation, but rather that of the conjugate minimal representation. Indeed, evaluating (\ref{anticharges1}) for 
${h}_1 =- 1$, ${h}_2=1$, ${h}_3=0$ with $\sigma_3{\psi}_0=0$ we find $\bar{\psi}_1 = 1$, $\bar{\psi}_2=1$ and $\bar{\psi}_3=1$, which reproduces indeed (\ref{unhattedbarxcharges}). 
\smallskip

Incidentally, the situation is precisely reverse with respect to the hatted modes. In that case, we need to evaluate the charges
\begin{eqnarray}
\hat{\psi}_2 & = & 2  \sum_{m\in\mathbb{Z}} \sum_i \, : \bar{\alpha}^i_{-m}  \alpha^i_{m}: \\
\hat{\psi}_3 & = & \sum_{m\in\mathbb{Z}} \sum_i \, (1-3m) \, : \bar{\alpha}^i_{-m} \alpha^i_{m} : \ . 
\end{eqnarray}
First consider the state $\bar\alpha^i_{-1}|0\rangle$ (that is relevant for the $\bar{x}_s$ modes), for which we find 
\be
\hat{{\psi}}_2 (\bar{x})= 2 \ , \qquad \hat{{\psi}}_3(\bar{x}) = -2 \ .
\ee
This is then of the form (\ref{charges}) with ${h}_1 =- 1$, ${h}_2=1$, ${h}_3=0$ and ${\s}_3 \hat{\psi}_0 = -1$; thus the $\bar{x}_r$ generators transform in the minimal representation with respect to the hatted modes. On the other hand, on the state $\alpha^i_{-1}|0\rangle$ that is relevant for the description of the $x_s$ modes, see eq.~(\ref{xdef}), the charges equal 
\be\label{chargesbos}
\hat\psi_2(x) = 2 \ , \qquad \hat\psi_3(x) = 4  \ . 
\ee
Now, this does not correspond to the minimal representation, i.e.\ it does not match (\ref{charges}) with ${h}_1 =- 1$, ${h}_2=1$, ${h}_3=0$ and ${\s}_3 \hat{\psi}_0 = -1$, but rather corresponds to the conjugate minimal representation, i.e., it agrees with (\ref{anticharges1}) for ${h}_1 =- 1$, ${h}_2=1$, ${h}_3=0$ with $\sigma_3\hat{\psi}_0=-1$. The situation for the $y$ and $\bar{y}$ generators is similar; we have summarized the representation properties of these generators in the Table~\ref{tab1}. 
\begin{table}
\begin{center}
\begin{tabular}{|c|c|c|}
\hline \\[-15pt]
\hbox{generator} & \hbox{unhatted modes ${\cal Y}$} & \hbox{hatted modes $\hat{{\cal Y}}$} \\ \hline
&& \\[-12pt]
$x$ & \hbox{minimal} & \hbox{conj minimal} \\
$\bar{x}$ &  \hbox{conj.\ minimal} & \hbox{minimal} \\
$y$ & \hbox{conj.\ minimal} & \hbox{minimal} \\
$\bar{y}$ &  \hbox{minimal} & \hbox{conj.\ minimal} \\ \hline
\end{tabular}
\end{center}
\caption{The representation properties of the fermionic generators.}\label{tab1}
\end{table}

\section{The Yangian at generic parameters} 

Next we want to make a proposal for how the algebra should be deformed away from the special point $\lambda=0$, see eq.~(\ref{hdef}). 
Our guiding principle is that, with respect to the the two bosonic affine algebras, denoted by $\mathcal{Y}$ and $\hat{\mathcal{Y}}$ respectively in the following, 
the fermionic generators sit in ``bi-minimal" representations.

\subsection{The generators in minimal representations}

Let us begin with studying the generators that transform in minimal (rather than conjugate minimal) representations. 
As we have seen above, the generator $x_s$ transforms in the minimal representation of $\mathcal{Y}$. 
By analogy with the construction of the bosonic affine Yangian, the operation of adding $x_s$ should therefore change the eigenvalue of the $\psi$ modes by --- this is $\psi(u)$ in eq.~(\ref{psiudef}) without the ``vacuum" factor $\psi_{0}(u) = (1 + \frac{\psi_0 \sigma_3}{u})$
\be\label{varphi2}
\varphi_2(u) = \frac{u(u +h_2)}{(u -h_1)(u -h_3)} \ . 
\ee
This then suggests that (\ref{psix}) should become 
\be
\psi(z) \, x(w)  \sim  \varphi_2(z-w) \, x(w) \psi(z) \ ,  \label{psixOPE} 
\ee
whose modes --- this can be deduced as in \cite{Gaberdiel:2017dbk}, see the discussion around eq.~(2.12) there ---  then satisfy
\be\label{psixgen}
[\psi_{r+2},{x}_s]-2[\psi_{r+1},{x}_{s+1}] +[\psi_r,{x}_{s+2 }] + h_2\Bigl([\psi_{r+1},{x}_{s }]-[\psi_r,{x}_{s+1 }]\Bigr)
+h_1 h_3 \psi_r x_s  =  0 \ . 
\ee
Note that this reduces to (\ref{psix}) for $h_2=1$ and $h_1h_3=0$. 
\smallskip

Before we proceed further, we can test this proposal by working out the charges of the state that is created by $x_{1/2}$ from the vacuum. 
Recall that on the vacuum state the bosonic and fermionic modes satisfy
\bal
e_i|0\>=\delta_{i,0} e_0|0\>\ , \qquad x_i |0\rangle = \delta_{i,1/2} \, x_{1/2} |0\rangle \ , \qquad
\bar{x}_i |0\rangle = \delta_{i,1/2} \, \bar{x}_{1/2} |0\rangle \ .
\eal
We want to confirm that the state $x_{1/2}|0\rangle$ has the charges of the minimal representation of $\mathcal{Y}$. 
We postulate that the initial condition (\ref{ini}) is now modified to 
\be\label{modinix}
[\psi_0,x_s]  = 0 \qquad \qquad  [\psi_{1},{x}_s]= - \frac{1}{h_2}\, {x}_s  \ .
\ee
Then it follows that 
\bal
\psi_0  \, x_{\frac{1}{2}} |0\>&=  N\, x_{\frac{1}{2}} |0\> \\
\psi_1 \, x_{\frac{1}{2}} |0\>&=  - \frac{1}{h_2} x_{\frac{1}{2}} |0\>  \ . 
\eal
In order to determine the higher charges we deduce from (\ref{psixgen}) that 
\bal
{}\psi_2 x_{1/2}  |0\rangle & = [\psi_2, x_{1/2}]  |0\rangle =  - h_2  [\psi_1,x_{1/2}] |0\rangle  - h_1 h_3 \psi_0 x_{1/2} |0\rangle \nonumber \\
& =  (1- h_1 h_3 \psi_0)\, x_{1/2} |0\rangle\\[8pt]
{}[\psi_3,x_{1/2}] |0\rangle & = [\psi_3, x_{1/2}]  |0\rangle =  - h_2  [\psi_2,x_{1/2}] |0\rangle - h_1 h_3 \psi_1 x_{1/2} |0\rangle \nonumber \\
& = - h_2 (1- h_1 h_3 \psi_0)\, x_{1/2} |0\rangle + \frac{h_1 h_3}{h_2}\, x_{1/2} |0\rangle \nonumber \\
& = \Bigl(\sigma_3 \psi_0  - \frac{h_2^2-h_1 h_3}{h_2} \Bigr) \, x_{1/2} |0\rangle \ .
\eal
Since $h_2=-(h_1+h_3)$, the last equation can now be rewritten as 
\be
\psi_3 \, x_{\frac{1}{2}} |0\>= \Bigl( \sigma_3 \psi_0 - \frac{h_1^2 + h_1 h_3 + h_3^2}{h_2}  \Bigr) x_{\frac{1}{2}} |0\> \ . 
\ee
These charges then agree precisely with eq.~(\ref{charges}), thus confirming that our ansatz (\ref{psixOPE}) leads to states with the correct charges. 

We also make the ansatz
\bal
e(z) \, x(w) & \sim  G(\Delta)\, x(w) \, e(z)  \label{g1} \\ 
f(z) \, x(w) & \sim  H(\Delta)\, x(w) \, f(z) \ , \label{g2}
\eal
where here and in the following 
\begin{equation}
\Delta\equiv z-w \ .
\end{equation}
$G(\Delta)$ and $H(\Delta)$ are functions that will be constrained further below, see Section~\ref{efOPE}. 

\subsubsection{Other minimal generators}

\begin{figure}[h!]
	\centering
	\includegraphics[trim=0cm 15cm 0cm 4cm, width=1.0\textwidth]{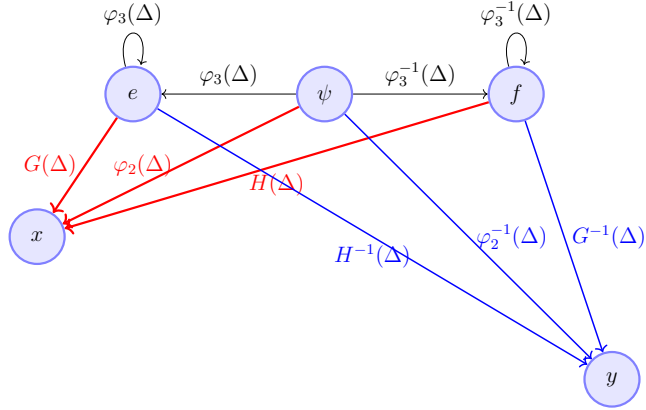}
	\caption{The OPE relations involving $x$ and $y$. 
	}
	\label{figOPEbosonic-xy}
	
\end{figure}

The construction works similarly for the generators $y_r$, which behave like the conjugate operators to $x_r$, i.e.\ they are like the $f_r$  modes relative to $e_r$ in the bosonic affine Yangian. 
Because of that we expect them to satisfy the inverse OPE, cf.\ eq.~(\ref{psiegen}) and (\ref{psifgen})
\be
\psi(z) \, {y}(w)  \sim  \varphi^{-1}_2(\Delta) \,\, {y}(w) \psi(z)  \ . \label{psiybOPEm} 
\ee
Given the simple relation between (\ref{psixOPE}) and (\ref{psiybOPEm}), we postulate that also the $y$-analogues of (\ref{g1}) and (\ref{g2}) only involve simple inverses. The structure of the OPEs of $x$ and $y$ with respect to the unhatted fields can then be summarized by the diagram of Figure~\ref{figOPEbosonic-xy}. 
 
\begin{figure}[h!]
	\centering
	\includegraphics[trim=0cm 15cm 0cm 4cm, width=1.0\textwidth]{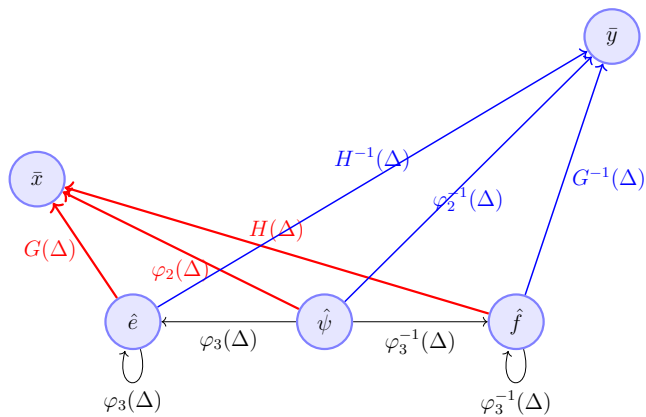}
	\caption{The OPE relations involving $\bar{x}$ and $\bar{y}$. 
	}
	\label{figOPEbosonicxy-conj}
\end{figure}

The analysis is completely analogous for the two fields $\bar{x}$ and $\bar{y}$ with respect to the hatted fields, and the structure of the corresponding OPEs can thus be similarly realized, see Figure~\ref{figOPEbosonicxy-conj}. 

\subsection{The generators in conjugate minimal representations}

\begin{figure}[h!]
	\centering
		\includegraphics[trim=2cm 12cm 0cm 4cm, width=1.0\textwidth]{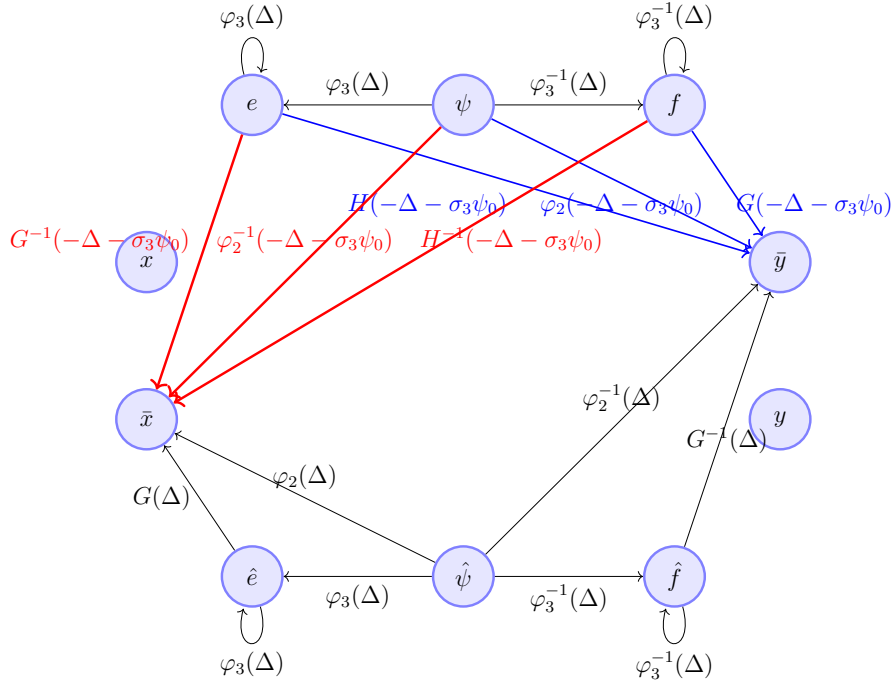}
	\caption{The additional relations (in red and blue) together with some of the relations from Figs.~\ref{figOPEbosonic-xy} and \ref{figOPEbosonicxy-conj} for comparison. 
	}
	\label{figOPEbosonicfull}
	\end{figure}

\begin{figure}[h!]
	\centering
	\includegraphics[trim=2cm 12cm 0cm 4cm, width=1.0\textwidth]{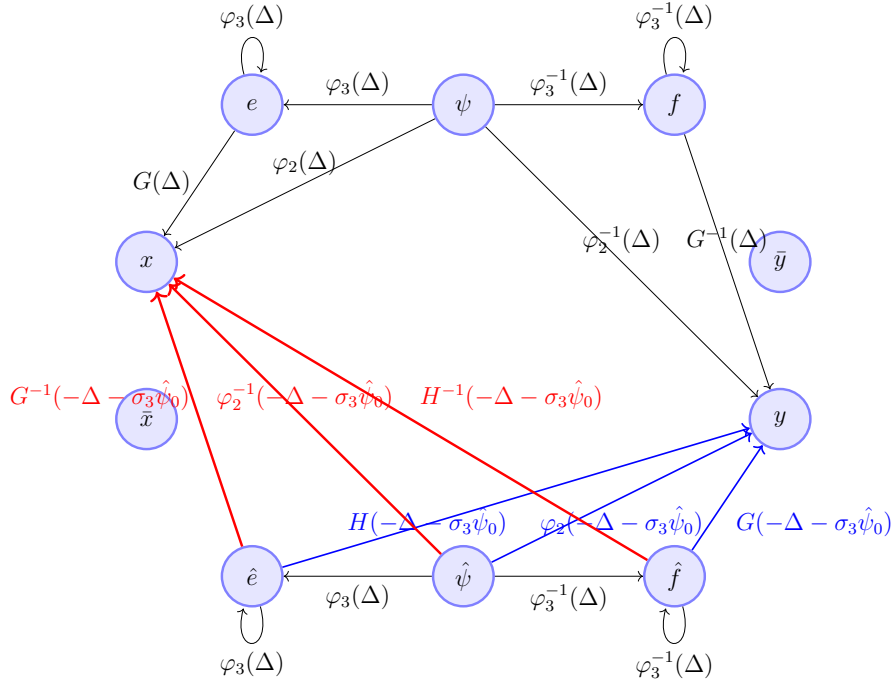}
	\caption{The additional relations (in red and blue) together with some of the relations from Figs.~\ref{figOPEbosonic-xy} and \ref{figOPEbosonicxy-conj} for comparison. 
	}
	\label{figOPEbosonicfull-conj}
	
\end{figure}

A more interesting case are the OPEs of the unhatted fields with $\bar{x}$ and $\bar{y}$, or equivalently, that of the hatted fields with $x$ and $y$. For concreteness, let us describe the former case in detail. 
Since the charges of the conjugate minimal representation are described by (\ref{conjchg}), the natural ansatz for the OPE is thus 
\be
\psi(z) \, \bar{x}(w)  \sim  \varphi^{-1}_2(-\Delta - \psi_0\sigma_3) \, \, \bar{x}(w) \psi(z) \ . \label{psixbOPE}  
\ee
By the usual arguments, this then leads to the commutation relations 
\bal
[\psi_{r+2},\bar{x}_s] & -2[\psi_{r+1},\bar{x}_{s+1}] +[\psi_r,\bar{x}_{s+2 }] + (- h_2+2\psi_0\sigma_3 ) \Bigl([\psi_{r+1},\bar{x}_{s }]-[\psi_r,\bar{x}_{s+1 }]\Bigr) \nonumber \\
& \qquad + \psi_0 \sigma_3 (\psi_0 \sigma_3 - h_2) [\psi_r,\bar{x}_s]  - h_1 h_3\, \bar{x}_s  \psi_r    =  0 \ , \label{hatpsixbgen}
\eal
which reduces indeed to (\ref{psibarx}) in the free field limit. Again, before proceeding further, we should check that this gives the correct charges on the corresponding states. In analogy to (\ref{modinix}) we now postulate 
\be
[\psi_0,\bar{x}_s]  =  0 \qquad \qquad  [\psi_{1},\bar{x}_s] = \frac{1}{h_2}  \bar{x}_s  \ . 
\ee
Then we find, using (\ref{hatpsixbgen})
\begin{eqnarray}
\psi_1 \, \bar{x}_{1/2} |0\rangle & = & \frac{1}{h_2} \bar{x}_{1/2} |0\rangle \\
\psi_2 \, \bar{x}_{1/2} |0\rangle  & = & 
[\psi_2,\bar{x}_{1/2}] \, |0\rangle = (h_2 - 2 \psi_0 \sigma_3) [\psi_1,\bar{x}_{1/2}] \, |0\rangle + h_1 h_3 \psi_0 \bar{x}_{1/2} \, |0\rangle \nonumber \\
& = & (1 - h_1 h_3 \psi_0) \bar{x}_{1/2} |0\rangle\\[6pt]
\psi_3 \, \bar{x}_{1/2} |0\rangle  & = & (h_2 - 2 \psi_0 \sigma_3) [\psi_2,\bar{x}_{1/2}] \, |0\rangle - \psi_0 \sigma_3 (\psi_0 \sigma_3 - h_2)  [\psi_1,\bar{x}_{1/2}]|0\rangle \nonumber \\
& = & \Bigl[ (h_2 - 2 \psi_0 \sigma_3)  (1 - h_1 h_3 \psi_0)  - \frac{\psi_0 \sigma_3 (\psi_0 \sigma_3 - h_2) }{h_2}  \Bigr] \bar{x}_{1/2} |0\rangle  \nonumber \\
& = & (1 - h_1 h_3 \psi_0) \bigl( h_2 - 2 \psi_0 \sigma_3 +\psi_0 \sigma_3)  \, \bar{x}_{1/2} |0\rangle \nonumber \\
& = & h_2 (1 - h_1 h_3 \psi_0)^2 \, \bar{x}_{1/2} |0\rangle  \ , 
\end{eqnarray}
thus giving the correct charges of the anti-minimal representation, see eq.~(\ref{anticharges1}). The structure of the corresponding OPEs can therefore be summarized as in Figure~\ref{figOPEbosonicfull}. The situation for the hatted fields with respect to $x$ and $y$ is completely analogous and summarized in Figure~\ref{figOPEbosonicfull-conj}.


\subsection{The OPEs with $e$ and $f$}\label{efOPE}

Since the $\psi$ field appears in the OPE of the $e$ and the $f$ field, see eq.~(\ref{efrel}),  
we can also deduce constraints on the OPE of the $e$ and $f$ field with $x$ from that with $\psi$. 
To this end, we recall the ansatz from eqs.~(\ref{g1}) and (\ref{g2})
\bal
e(z) \, x(w) & \sim  G(\Delta)\, x(w) \, e(z)  \label{g1p} \\ 
f(z) \, x(w) & \sim  H(\Delta)\, x(w) \, f(z) \ . \label{g2p}
\eal
Note that, just like the identities of eqs.~(\ref{eegen}) -- (\ref{psifgen}), these relations cannot be exactly correct, but are only true up to terms that are regular at either $z=0$ or $w=0$, see the discussion around eq.~(5.15) in \cite{Gaberdiel:2017dbk}. 
Applying this identity twice we find that 
\bal
e(z_1)\, f(z_2)  \, x(w) & \sim  G(z_1-w)\, H(z_2-w)\, x(w) \, e(z_1)  f(z_2)  \label{52}\\ 
f(z_2) \, e(z_1) \, x(w) & \sim  G(z_1-w)\, H(z_2-w)\, x(w) \, f(z_2)\, e(z_1) \ . \label{53}
\eal
Subtracting the two equations from one another and using (\ref{efrel}) we thus deduce that 
\be
\frac{ \psi(z_1)  x(w) - \psi(z_2) x(w) }{z_1-z_2}  \sim G(z_1-w)\, H(z_2-w)\, 
\frac{ x(w)\, \psi(z_1)  - x(w)\, \psi(z_2) }{z_1-z_2}  \ . 
\ee
Next we apply (\ref{psixOPE}) to the left-hand-side, from which we deduce that this equals
\be
\frac{ \varphi_2(z_1-w) x(w)\, \psi(z_1)  - \varphi_2(z_2-w) x(w)\, \psi(z_2) }{z_1-z_2}  \ .
\ee
Thus it follows that 
\be
\frac{ \varphi_2(z_1-w)}{z_1-z_2} \sim \frac{ \varphi_2(z_2-w)}{z_1-z_2}  \sim \frac{G(z_1-w)\, H(z_2-w)}{z_1-z_2} \ .
\ee
Because these identities are only true up to regular terms, this implies that the functions $G(\Delta)$ and $H(\Delta)$ have to satisfy the identity
\be
\boxed{G(\Delta) \, H(\Delta) = \varphi_2(\Delta) \ .} 
\ee
The most natural ansatz that is compatible with the free field limit (see below) is 
\be\label{GHdef}
G(\Delta) = \frac{\Delta+h_2}{\Delta-h_1} \ , \qquad
H(\Delta) = \frac{\Delta}{\Delta-h_3} \ ,
\ee
In terms of commutators this leads to 
\bal
0 & = [e_{r+1},x_s]  - [e_r,x_{s+1}] - h_1 [e_r,x_s] + h_3 x_s\, e_r   \label{ex2}\\
0 & = [f_{r+1},x_s] - [f_r,x_{s+1}] - h_3 f_r x_s \ , 
\eal
which reduces indeed to the correct free field answers, see eqs.~(\ref{exmain}) and (\ref{fxmain}). However, this ansatz cannot be right since there are two box-descendants of the state generated by $x_{1/2}|0\rangle$ --- this follows from the bosonic structure of the minimal representation --- and hence the function $G(\Delta)$ must have two poles \cite{GLP}. In fact, one can use the representation theory to constrain the function $G(\Delta)$ (and hence $H(\Delta)$) further, but this goes beyond the scope of the present paper and will be described elsewhere \cite{GLP}.

\subsection{The ${\cal N}=2$ algebra}

Our starting point in Section~\ref{sec:2} was the ${\cal W}^{({\cal N}=2)}_{\infty}[\lambda]$ algebra, and we can now try to identify its generators with those of the supersymmetric affine Yangian. The vacuum character, see eq.~(\ref{chi0}), contains two spin $s=1$ fields: the $\mathfrak{u}(1)$ generator of the ${\cal N}=2$ superconformal algebra, as well as the extra bosonic generator that we added by hand to the ${\cal W}^{({\cal N}=2)}_{\infty}[\lambda]$ vacuum character (\ref{chi0p}). 
This free boson should be completely decoupled, and its zero mode be identified with the central generator
\be\label{U0}
U_0 \equiv \psi_1 + \hat{\psi}_1 \ . 
\ee
Obviously, $U_0$ commutes with all $e_r$, $f_r$, $\hat{e}_r$ and $\hat{f}_r$ generators, and because of the relations we have imposed, it also commutes with the $x_s$ and $\bar{x}_s$ generators.

We also know that the total M\"obius generators correspond to 
\be\label{Moebius}
L_{-1} = e_1 + \hat{e}_1 \ , \qquad L_1 = - f_1 - \hat{f}_1 \ , \qquad L_0 = \frac{1}{2} \bigl( \psi_2 + \hat{\psi}_2 \bigr) \ ,
\ee
and it is thus natural to assume that the $\pm 1$ modes of the decoupled boson are 
\be
U_{-1} = e_0 + \hat{e}_0 \ , \qquad U_1 = - \bigl( f_0  + \hat{f}_0 \bigr) \ . 
\ee
The $\mathfrak{u}(1)$ generator of the ${\cal N}=2$ algebra, on the other hand, should be identified with 
\be
J_0 = (\sigma_3\hat{\psi}_0) \, \psi_1 - (\sigma_3{\psi}_0) \,\hat{\psi}_1  \ , 
\ee
and
\be
J_{-1} = (\sigma_3\hat{\psi}_0)\,  e_0 - (\sigma_3{\psi}_0)\, \hat{e}_0 \ , \qquad J_1 = -(\sigma_3\hat{\psi}_0)\, f_0+ (\sigma_3{\psi}_0)\, \hat{f}_0 \ . 
\ee
Then $[J_m,U_n]=0$ for $m,n\in\{0,\pm 1\}$, and we find 
\be
[J_m,J_n ] = \frac{c^{({\cal N}=2)}}{3}\, m \delta_{m,-n} \ , \qquad \hbox{with} \qquad 
c^{({\cal N}=2)} =   3 \, (\sigma_3)^2 \psi_0\, \hat\psi_0\,  ( \psi_0+ \hat\psi_0 ) \ . 
\ee
In terms of the dictionary of \cite{Gaberdiel:2017dbk}, this central charge then equals 
\be
c^{({\cal N}=2)} = \frac{3Nk}{N+k+1} \ , 
\ee 
in agreement with the ${\cal N}=2$ central charge, see eq.~(\ref{cNkN=2}). The other bosonic generators can be similarly identified: for each integer $s$, there are two decoupled bosonic fields, see the discussion below (\ref{assumption}), and they can be identified with the affine Yangian generators of the two bosonic affine Yangians, using the dictionary of \cite{Gaberdiel:2017dbk}.

This leaves us with the fermionic generators. The lowest fermionic generators are the supercharge generators $G^\pm_{r}$, which, at the free point, can be identified with 
\be\label{iden}
x_{1/2} = \frac{1}{\sqrt{2}} \,  G^+_{-3/2} \ , \qquad  \bar{x}_{1/2} = \frac{1}{\sqrt{2}} \, G^-_{-3/2} \ .
\ee
It would be natural to postulate this identification also for generic $h_i$. 
However, there is a problem with this proposal. 
The ${\cal N}=2$ generators should commute with the decoupled free boson described by $U_n$. 
But even at the free point one finds 
\be\label{Ue0}
[e_0 + \hat{e}_0, x_{1/2}] = \sum_{n\neq 0} \sum_{i} \frac{1}{n} \, : \bar{\chi}^i_{-5/2-n} \alpha^i_n : \ .
\ee
The origin of this problem is that the $\hat{e}_0$ generator corresponds to the $-1$ mode of a non-local field, 
\be
\hat{e}_0 = - \sum_{m\neq 0} \sum_i \, \frac{1}{m} \,  : \bar{\alpha}^i_{-m-1} \alpha^i_m : \ ,
\ee
see also the discussion below eq.~(\ref{hated3}). The fact that $\hat{e}_0$ is non-local at the free point is an artefact of the free limit --- for generic $\lambda$, both $e_0$ and $\hat{e}_0$ describe the $-1$ mode of local fields, as follows from the discussion in \cite{Gaberdiel:2017dbk}. One may therefore suspect that the fact that $[U_0,x_{1/2}]\neq 0$ is purely a free-field artefact, but this is not the case: as we show in Appendix~\ref{app:supercharge}, with the above identifications, this problem persists for generic $\lambda$. 

We believe that the resolution of this problem is that we need to correct the identification (\ref{iden}) by (non-local) correction terms. The fact that such non-local correction terms appear is maybe not surprising in view of the fact that also in the bosonic setting non-local correction terms were required for the identification of the spin $3$ and $4$ fields, see  \cite{Gaberdiel:2017dbk}. The relevant analysis is, however, rather cumbersome, and we leave it to future work.

\section{Conclusions}

In this paper we have found some of the defining relations of the Yangian algebra that is expected to be isomorphic to ${\cal W}^{({\cal N}=2)}_{\infty}[\lambda]$, the ${\cal N}=2$ superconformal version of ${\cal W}_\infty[\lambda]$. 
We have extensively used the fact that ${\cal W}^{({\cal N}=2)}_{\infty}[\lambda]$ contains two commuting bosonic ${\cal W}_{\infty}[\mu]$ algebras, each of which in turn is isomorphic to an affine Yangian of $\mathfrak{gl}_1$. 
The additional generators transform in bi-minimal representations with respect to these two ${\cal W}_\infty[\mu]$ algebras. 
We have shown how this translates into explicit commutation relations for the additional Yangian generators --- our main technical advance is the description of the conjugate representations, see Section~4.2. 
This has allowed us to make a proposal for at least some of the defining relations of the ${\cal N}=2$ superconformal affine Yangian. We have also checked --- in fact this was an important guiding principle --- that these relations reduce to the expected identities in the free field case ($\lambda=0$). 

There are many open questions which we hope to address in the future. First of all, it would be nice to construct the representation theory of this Yangian algebra, see \cite{GLP} for first steps in this direction; this will involve two plane partitions on which the various generators should have some natural action. (The two affine Yangians of $\mathfrak{gl}_1$ act separately on each, while the additional bi-minimal generators generate infinite rows of boxes (and anti-boxes), connecting the two plane partitions.) Among other things, this would allow us to prove the consistency of our construction and to find the remaining relations. 
It would also be interesting to establish the dictionary to the ${\cal W}^{({\cal N}=2)}_{\infty}[\lambda]$ algebra in more detail, generalizing the construction of \cite{Gaberdiel:2017dbk} to the current context, and to explore the various duality symmetries this picture suggests. Note that the construction selects out one of the three directions of each plane partition, thereby breaking the $S_3$ symmetry  \cite{Gaberdiel:2012ku}  of each affine Yangian of $\mathfrak{gl}_1$ to a $\mathbb{Z}_2$ symmetry that exchanges the remaining two directions. Together with the exchange symmetry of the two affine Yangians, the ${\cal N}=2$ affine Yangian therefore has a $\mathbb{Z}_2\times \mathbb{Z}_2$ symmetry. We hope to come back to these questions in the near future.

\section*{Acknowledgements} 
We thank Tomas Prochazka and Miroslav Rapcak for sending us \cite{PR} prior to publication, and Tomas Prochazka for comments on the draft. We also thank Yang Lei for help with drawing the figures. The work of MRG is (partly) supported by the NCCR SwissMAP, funded by the Swiss National Science Foundation. WL is grateful for support from the ``Thousand talents grant" and from the ``Max Planck Partnergruppe". The work of CP is supported by the US Department of Energy under contract DE-SC0010010 Task A. HZ is partially supported by the General Financial Grant from the China Postdoctoral Science Foundation, with Grant No. 2017M611009. MRG thanks Beijing University for hospitality during the very final stages of this work. We gratefully acknowledge the hospitality of the Galileo Galilei Institute for Theoretical Physics (GGI) for hospitality, and INFN for partial financial support during the program ``New Developments in AdS3/CFT2 Holography".

\appendix

\section{Additional relations of the free field theory}\label{app:Serre}

In addition to the free field relations  that were given in the main body of the text, see eqs.~(\ref{psix}) -- (\ref{ini}), the commutation relations of $y_s$ and $\bar{y}_s$ are 
\bal
0 & = [\psi_{r+2},{y}_s]-2[\psi_{r+1},{y}_{s+1}] +[\psi_r,{y}_{s+2 }]+[\psi_{r+1},{y}_{s }]-[\psi_r,{y}_{s+1 }] \\
0 & = [{e}_{r+1 },  y_s] -  [{e}_r, y_{s +1}]  \\
0 & = [{f}_{r+1}, y_s]-[{f}_r, y_{s+1}] + [{f}_r, y_{s }] \\
0 & = [\psi_{r+2},\bar {y}_s]-2[\psi_{r+1},\bar {y}_{s+1}] +[\psi_r,\bar {y}_{s+2 }]-[\psi_{r+1},\bar {y}_{s }]+[\psi_r,\bar {y}_{s+1 }] \\
0 & = [{e}_{r+1 },\bar y_s] -  [{e}_r,\bar y_{s+1 }]  \\ 
0 & = [f_{r+1},\bar y_s]-[f_r,\bar y_{s+1}] - [f_r,\bar y_{s }]  \ ,
\eal
as well as 
\bal
0 & = [\hat\psi_{r+2},{y}_s]-2[\hat\psi_{r+1},{y}_{s+1}] +[\hat\psi_r,{y}_{s+2 }]-3[\hat\psi_{r+1},{y}_{s }]+3[\hat\psi_r,{y}_{s+1 }]+2 [\hat{\psi}_r,{y}_s] \nonumber \\ %
0 & = [\hat{e}_{r+1 },  y_s] -  [\hat{e}_r, y_{s +1}] -  [\hat{e}_r, y_{s }] \\
0 & = [\hat{f}_{r+1}, y_s]-[\hat{f}_r, y_{s+1}] - 2[\hat{f}_r, y_{s }]  \\
0 & = [\hat\psi_{r+2},\bar {y}_s]-2[\hat\psi_{r+1},\bar {y}_{s+1}] +[\hat\psi_r,\bar {y}_{s+2 }]+ [\hat\psi_{r+1},\bar {y}_{s }]- [\hat\psi_r,\bar {y}_{s+1 }] \\
0 & = [\hat{e}_{r+1 },\bar {y}_s] - [\hat{e}_r,\bar {y}_{s+1 }] \\
0 & = [\hat{f}_{r+1},\bar y_s]-[\hat{f}_r,\bar y_{s+1}] + [\hat{f}_r,\bar y_{s }] \ . 
\eal
The $P_r$ modes that were defined in eq.~(\ref{Pdef}), and their corresponding barred version
\be
\{ x_r, y_s\} = P_{r+s}  \ , \qquad \{\bar{x}_r,\bar{y}_s\} =\bar{P}_{r+s}  \ ,
\ee
are given explicitly by 
\be
P_1 =  \frac{1}{2} \bigl( \psi_2 + \hat{\psi}_2) + \frac{3}{2} \psi_1 + N \ , \qquad
\bar{P}_1 =  \frac{1}{2} \bigl( \psi_2 + \hat{\psi}_2) - \frac{3}{2} \psi_1 + N \ , 
\ee
as well as (for $r\geq 2$)
\bal
P_r & =  {  \sum_{m\in \mathbb{Z}+\frac{1}{2}} \sum_{i} (m+\tfrac{3}{2})(m+\tfrac{1}{2})^{r-1} \, :{\chi}^i_{-m} \bar{\chi}^i_{m}:
+\sum_{u\in \mathbb{Z}}  \sum_{i}  \bigl(u-1 \bigr)^{r-1} \, : \alpha^i_{-u} \bar\alpha^i_{u}:}   \nonumber \\
\bar P_r & =  {  \sum_{m\in \mathbb{Z}+\frac{1}{2}} \sum_{i} (m-\tfrac{3}{2})(m-\tfrac{1}{2})^{r-1} \, :{\chi}^i_{-m}\bar{\chi}^i_{m} :
+\sum_{u\in \mathbb{Z}}  \sum_{i}  \bigl(u+1 \bigr)^{r-1} \, : \alpha^i_{-u} \bar\alpha^i_{u}:}  \ . \nonumber
\eal
They satisfy 
\bal
0&=[P_{i+2},\bar{x}_{j}]-2[P_{i+1},\bar{x}_{j+1}]+[P_{i},\bar{x}_{j+2}] + [P_{i+1},\bar{x}_{j}] -[P_{i},\bar{x}_{j+1}] - 2 [P_i,\bar{x}_j] \label{Pxb} \\ 
0&=[P_{i+2},\bar y_{j}]-2[P_{i+1},\bar y_{j+1}]+[P_{i},\bar y_{j+2}] + [P_{i+1},\bar y_{j}] -[P_{i},\bar y_{j+1}] - 2 [P_i,\bar y_j] \label{Pyb}\\ 
0&=[P_{i+2}, e_{j}]-2[P_{i+1}, e_{j+1}]+[P_{i}, e_{j+2}] -([P_{i+1}, e_{j}] -[P_{i}, e_{j+1}]) \label{Pe}\\
0&=[P_{i+2}, f_{j}]-2[P_{i+1}, f_{j+1}]+[P_{i}, f_{j+2}] -([P_{i+1}, f_{j}] -[P_{i}, f_{j+1}]) \label{Pf}\\
0&=[P_{i+2},\hat e_{j}]-2[P_{i+1},\hat e_{j+1}]+[P_{i},\hat e_{j+2}] + 3([P_{i+1},\hat e_{j}] -[P_{i},\hat e_{j+1}]) + 2 [P_i,\hat e_j]  \label{Phate}\\
0&=[P_{i+2},\hat f_{j}]-2[P_{i+1},\hat f_{j+1}]+[P_{i},\hat f_{j+2}] + 3([P_{i+1},\hat f_{j}] -[P_{i},\hat f_{j+1}]) + 2 [P_i,\hat f_j] \label{Phatf}
\eal
and 
\bal
0&=[\bar  P_{i+2},{x}_{j}]-2[\bar P_{i+1},{x}_{j+1}]+[\bar P_{i},{x}_{j+2}] - [\bar P_{i+1},{x}_{j}] +[\bar P_{i},{x}_{j+1}] - 2 [\bar P_i,{x}_j]  \label{Pbx}\\
0&=[ \bar{P}_{i+2}, y_{j}]-2[\bar P_{i+1}, y_{j+1}]+[\bar P_{i}, y_{j+2}] - [\bar P_{i+1}, y_{j}] +[\bar P_{i}, y_{j+1}] - 2 [\bar P_i, y_j] \label{Pby}\\ 
0&=[\bar P_{i+2}, \hat e_{j}]-2[\bar P_{i+1}, \hat e_{j+1}]+[\bar P_{i}, \hat e_{j+2}] -([\bar P_{i+1},\hat  e_{j}] -[\bar P_{i}, \hat e_{j+1}]) \label{Pbe}\\
0&=[\bar P_{i+2}, \hat f_{j}]-2[\bar P_{i+1}, \hat f_{j+1}]+[\bar P_{i}, \hat f_{j+2}] -([\bar P_{i+1}, \hat f_{j}] -[\bar P_{i}, \hat f_{j+1}]) \label{Pbf}\\
0&=[\bar P_{i+2}, e_{j}]-2[\bar P_{i+1}, e_{j+1}]+[\bar P_{i}, e_{j+2}] + ([\bar P_{i+1}, e_{j}] -[\bar P_{i}, e_{j+1}]) \label{Pbhate}\\
0&=[\bar P_{i+2}, f_{j}]-2[\bar P_{i+1}, f_{j+1}]+[\bar P_{i}, f_{j+2}] + ([\bar P_{i+1},f_{j}] -[\bar P_{i}, f_{j+1}])  \ . \label{Pbhatf}
\eal
\smallskip

\section{The defining relations}\label{app:rels}

In this appendix we collect some of the defining relations of the supersymmetric affine Yangian.

\subsection{The OPE like description}

\bal
\psi(z) \, x(w) & \sim  \varphi_2(\Delta) \, x(w) \psi(z)  \label{psixOPEa} \\
\hat\psi(z) \, x(w)  &\sim   \varphi^{-1}_2(-\Delta - \hat\psi_0\sigma_3) \, x(w) \hat\psi(z)   \label{hatpsixOPE} \\
\psi(z) \, \bar{x}(w)  &\sim   \varphi^{-1}_2(-\Delta - \psi_0\sigma_3) \, \bar{x}(w) \psi(z)  \label{psixbOPE} \\
\hat\psi(z) \, \bar{x}(w)  &\sim  \varphi_2(\Delta) \, \bar{x}(w) \hat\psi(z) \ , \label{hatpsixbOPE} 
\eal
where $\varphi_2(u)$ is defined in eq.~(\ref{varphi2}). For the $y$ fields we find 
\bal
\psi(z) \, {y}(w)  &\sim  \varphi^{-1}_2(\Delta) \, {y}(w) \psi(z)  \label{psiybOPE} \\
\hat\psi(z) \, {y}(w)  &\sim  \varphi_2(-\Delta - \hat\psi_0\sigma_3) \, {y}(w) \hat\psi(z)  \label{hatpsiybOPE} \\
\psi(z) \,\bar  y(w)  &\sim   \varphi_2(-\Delta - \psi_0\sigma_3) \, \bar y(w) \hat\psi(z)   \label{psiyOPE} \\
\hat\psi(z) \, \bar  y(w)  &\sim   \varphi^{-1}_2(\Delta) \, \bar  y(w) \hat\psi(z)   \label{hatpsiyOPE} \ .
\eal

\subsection{The mode relations}\label{app:B2}

In terms of modes, these identities are 
\bal
[\psi_{r+2},{x}_s]& -2[\psi_{r+1},{x}_{s+1}] +[\psi_r,{x}_{s+2 }] + h_2\Bigl([\psi_{r+1},{x}_{s }]-[\psi_r,{x}_{s+1 }]\Bigr)
+h_1 h_3 \psi_r x_s  =  0  \nonumber \\
[\hat\psi_{r+2},{x}_s] & -2[\hat\psi_{r+1},{x}_{s+1}] +[\hat\psi_r,{x}_{s+2 }] - (h_2-2\hat{\psi}_0\sigma_3 ) \Bigl([\hat\psi_{r+1},{x}_{s }]-[\hat\psi_r,{x}_{s+1 }]\Bigr) \nonumber \\
& \qquad + (h_1+\hat{\psi}_0\sigma_3 )(h_3+\hat{\psi}_0\sigma_3 ) [\hat\psi_r,{x}_s]  - h_1 h_3 \,  \hat\psi_r {x}_s   =  0   \nonumber  \\[4pt]
[\psi_{r+2},\bar{x}_s] & -2[\psi_{r+1},\bar{x}_{s+1}] +[\psi_r,\bar{x}_{s+2 }] + (-h_2+2\psi_0\sigma_3 ) \Bigl([\psi_{r+1},\bar{x}_{s }]-[\psi_r,\bar{x}_{s+1 }]\Bigr) \nonumber \\
& \qquad + \psi_0 \sigma_3 (\psi_0 \sigma_3 - h_2) [\psi_r,\bar{x}_s]  - h_1 h_3\, \bar{x}_s  \psi_r    =  0 \nonumber  \\[4pt]%
[\hat\psi_{r+2},\bar{x}_s] & -2[\hat\psi_{r+1},\bar{x}_{s+1}] +[\hat\psi_r,\bar{x}_{s+2 }] + h_2\Bigl([\hat\psi_{r+1},\bar{x}_{s }]-[\hat\psi_r,\bar{x}_{s+1 }]\Bigr)
+h_1 h_3 \hat\psi_r \bar{x}_s  =  0 \nonumber \\
[\psi_{r+2},{y}_s] & -2[\psi_{r+1},{y}_{s+1}] +[\psi_r,{y}_{s+2 }] +h_2\Bigl([\psi_{r+1},{y}_{s }]-[\psi_r,{y}_{s+1 }]\Bigr) \nonumber \\
& \qquad   - h_1 h_3\,    {y}_s \psi_r  =  0 \nonumber   
\eal
\bal 
[\hat\psi_{r+2},{y}_s] & -2[\hat\psi_{r+1},{y}_{s+1}] +[\hat\psi_r,{y}_{s+2 }] - (h_2-2\hat\psi_0\sigma_3 ) \Bigl([\hat\psi_{r+1},{y}_{s }]-[\hat\psi_r,{y}_{s+1 }]\Bigr) \nonumber \\
& \qquad + \hat\psi_0 \sigma_3 (\hat\psi_0 \sigma_3 - h_2) [\hat\psi_r,{y}_s]  + h_1 h_3\,   \hat\psi_r {y}_s   =  0   \nonumber  \\[4pt]
[\psi_{r+2},\bar {y}_s]& -2[\psi_{r+1},\bar {y}_{s+1}] +[\psi_r,\bar {y}_{s+2 }]  - (h_2-2\psi_0\sigma_3 )\Bigl([\psi_{r+1},\bar {y}_{s }]-[\psi_r,\bar {y}_{s+1 }]\Bigr) \nonumber  \\
&\qquad + \psi_0 \sigma_3 (\psi_0 \sigma_3 - h_2) [\psi_r,\bar {y}_s] +h_1 h_3  \psi_r \bar y_s  =  0 \nonumber   \\[4pt]
[\hat\psi_{r+2},\bar {y}_s]& -2[\hat\psi_{r+1},\bar {y}_{s+1}] +[\hat\psi_r,\bar {y}_{s+2 }] + h_2\Bigl([\hat\psi_{r+1},\bar {y}_{s }]-[\hat\psi_r,\bar {y}_{s+1 }]\Bigr)
+h_1 h_3 \bar y_s  \hat\psi_r =  0  \ .\nonumber 
\eal
For completeness, we also give the mode relations of the bosonic generators  \cite{Prochazka:2015deb,Gaberdiel:2017dbk}, which we write only for the unhatted generators (but which hold similarly also for the hatted generators).
\begin{eqnarray}
0 & = & [\psi_j,\psi_k] \label{Y0} \\
\psi_{j+k} & = & [e_j,f_k] \label{Y3} 
\end{eqnarray}
\begin{eqnarray}
\sigma_3 \{e_j,e_k \}  & = & [e_{j+3},e_k] - 3  [e_{j+2},e_{k+1}] + 3  [e_{j+1},e_{k+2}]  - [e_{j},e_{k+3}]   \nonumber \\
& & \ + \sigma_2 [e_{j+1},e_{k}] - \sigma_2  [e_{j},e_{k+1}]  \ ,\label{Y1}  \\
-  \sigma_3 \{f_j,f_k \} & = &  [f_{j+3},f_k] - 3  [f_{j+2},f_{k+1}] + 3  [f_{j+1},f_{k+2}]  - [f_{j},f_{k+3}]  \nonumber \\
& & \ + \sigma_2 [f_{j+1},f_{k}] - \sigma_2  [f_{j},f_{k+1}]   \label{Y2} \\
\sigma_3 \{\psi_j,e_k \}  & = &  [\psi_{j+3},e_k] - 3  [\psi_{j+2},e_{k+1}] + 3  [\psi_{j+1},e_{k+2}]  - [\psi_{j},e_{k+3}]  \nonumber \\
& & \ + \sigma_2 [\psi_{j+1},e_{k}] - \sigma_2  [\psi_{j},e_{k+1}]  \label{Y4} \\
- \sigma_3 \{\psi_j,f_k \}  & = &  [\psi_{j+3},f_k] - 3  [\psi_{j+2},f_{k+1}] + 3  [\psi_{j+1},f_{k+2}]  - [\psi_{j},f_{k+3}]  \nonumber \\
& & \ + \sigma_2 [\psi_{j+1},f_{k}] - \sigma_2  [\psi_{j},f_{k+1}]  \label{Y5} \ .
\end{eqnarray}
In addition, they satisfy the Serre relations
\be\label{Serre}
{\rm Sym}_{(j_1,j_2,j_3)} [e_{j_1},[e_{j_2},e_{j_3+1}]] = 0 \ , \qquad
{\rm Sym}_{(j_1,j_2,j_3)} [f_{j_1},[f_{j_2},f_{j_3+1}]] = 0 \ . 
\ee

\subsection{The initial conditions}

The modified initial conditions, generalizing (\ref{ini}) are 
\be\label{inigen}
\begin{array}{rclrcl}
{} [\psi_0,x_s] & = & 0 \qquad \qquad & [\psi_{1},{x}_s]&=& - h_2^{-1} \, {x}_s  \\
{}[\hat\psi_{0},{x}_s]&=& 0  \qquad \qquad & [\hat\psi_{1},{x}_s]&=& h_2^{-1} x_s\\
{} [\psi_0,\bar{x}_s] & = & 0 \qquad \qquad & [\psi_{1},\bar{x}_s]&=&  h_2^{-1} \, \bar{x}_s  \\
{}[\hat\psi_{0},\bar{x}_s]&=& 0  \qquad \qquad & [\hat\psi_{1},\bar{x}_s]&=&- h_2^{-1} \bar{x}_s \ .
\end{array}
\ee
Furthermore, 
\be\label{inigen}
\begin{array}{rclrcl}
	{} [\psi_0,\bar y_s] & = & 0 \qquad \qquad & [\psi_{1},\bar {y}_s]&=& - h_2^{-1} \, \bar {y}_s  \\
	{}[\hat\psi_{0},\bar {y}_s]&=& 0  \qquad \qquad & [\hat\psi_{1},\bar {y}_s]&=& h_2^{-1} \bar y_s\\
	{} [\psi_0,{y}_s] & = & 0 \qquad \qquad & [\psi_{1},{y}_s]&=&  h_2^{-1} \, {y}_s  \\
	{}[\hat\psi_{0},{y}_s]&=& 0  \qquad \qquad & [\hat\psi_{1},{y}_s]&=&- h_2^{-1} {y}_s \ .
	\end{array}
\ee
Finally, the initial relations of the bosonic generators are 
\be\label{ini1}
[\psi_0,e_j] = 0 \ , \qquad [\psi_1,e_j] = 0 \ , \qquad [\psi_2,e_j] = 2 e_j  \ ,
\ee
and
\be\label{ini2}
[\psi_0,f_j] = 0 \ , \qquad [\psi_1,f_j] = 0 \ , \qquad [\psi_2,f_j] = - 2 f_j  \ .
\ee

\section{Supercharge constraints}\label{app:supercharge}

We start by assuming that the decoupled $\mathfrak{u}(1)$ generator commutes with the $x_{1/2}$ generator, which we would like to identify with the supercharge, see eq.~(\ref{iden}), 
\be\label{basic1}
{} [ e_0 + \hat{e}_0,x_{1/2}] = 0 \ . 
\ee
We will deduce two identities from this assumption, using the commutation relations we have postulated. 
First we consider
\begin{equation}
\begin{aligned}
& [ [ (\psi_2 -\hat{\psi}_2),x_{1/2}], (e_0 + \hat{e}_0)] \\
& \qquad =  [ [ (\psi_2 -\hat{\psi}_2),(e_0 + \hat{e}_0)], x_{1/2} ]  + [ [(e_0 + \hat{e}_0), x_{1/2} ], (\psi_2 -\hat{\psi}_2)] \nonumber \\
& \qquad =  2 [ (e_0 - \hat{e}_0), x_{1/2} ]  = 4 [ e_0 , x_{1/2} ] \ .  \label{step1}
\end{aligned}
\end{equation}
For the left-hand side we then use 
\begin{eqnarray}
{}[\psi_2,x_{1/2}] & = & - \frac{2}{h_2} x_{3/2} + (1-h_1 h_3 \psi_0) \, x_{1/2} \\
{}[\hat{\psi}_2,x_{1/2}] & = &  \frac{2}{h_2} x_{3/2} + (1-h_1 h_3 \hat{\psi}_0) \, x_{1/2}
\end{eqnarray}
from which it follows, using again (\ref{basic1}), that 
\be\label{id1}
[x_{3/2},e_0+\hat{e}_0] = h_2 [x_{1/2},e_0] \ . 
\ee

The other identity can be derived similarly, except that now we consider the commutator with $\psi_3  + \hat{\psi}_3$, 
\begin{equation}
\begin{aligned}
&[ [ (\psi_3 +\hat{\psi}_3),x_{1/2}], (e_0 + \hat{e}_0)] 
\\ & \qquad =  [ [ (\psi_3 +\hat{\psi}_3),(e_0 + \hat{e}_0)], x_{1/2} ]  + [ [(e_0 + \hat{e}_0)], x_{1/2} ], (\psi_3 +\hat{\psi}_3)] \nonumber \\
& \qquad =  [ [( \psi_3 +\hat{\psi}_3),e_0 + \hat{e}_0], x_{1/2} ] \ .  \label{step2}
\end{aligned}
\end{equation}
Now the relevant charge relations are 
\be
[ \psi_3 +\hat{\psi}_3,x_{1/2}] = (10 + 2 h_1 h_3 \psi_0) x_{3/2} + d\, x_{1/2} 
\ee
where $d$ is some constant, and 
\be
[ \psi_3 +\hat{\psi}_3,e_0+\hat{e}_0] = 6 (e_1 + \hat{e}_1) + 2 \sigma_3 \psi_0\, e_0 + 2 \sigma_3 \hat{\psi}_0 \,\hat{e}_0 \ . 
\ee
This then leads to the identity 
\begin{eqnarray}
&&(10 + 2 h_1 h_3 \psi_0) [x_{3/2},e_0+\hat{e}_0]\nonumber\\ 
&& \qquad =  6 [e_1+\hat{e}_1,x_{1/2}] + 2 \sigma_3 \psi_0 [e_0,x_{1/2}] + 2 \sigma_3 \hat{\psi}_0 [\hat{e}_0,x_{1/2}] \nonumber  \\
&&\qquad =  6 [ \sigma_3 \hat{\psi}_0\, e_0 - \sigma_3 \psi_0\, \hat{e}_0,x_{1/2}] + 2 \sigma_3 \psi_0 [e_0,x_{1/2}] + 2 \sigma_3 \hat{\psi}_0 [\hat{e}_0,x_{1/2}] \nonumber\\
&&\qquad  =  h_2 (4  - 4 h_1 h_3 \psi_0) \, [x_{1/2},e_0] \ , \label{id2} 
\end{eqnarray}
where we have also used that 
\be
[e_1 + \hat{e}_1,x_{1/2}] = [L_{-1},x_{1/2}] = [ J_{-1},x_{1/2}] = [\sigma_3 \hat{\psi}_0 \, e_0 - \sigma_3 \psi_0\, \hat{e}_0,x_{1/2}]  \ ,
\ee
as follows from the fact that $x_{1/2} \sim G^+_{-3/2}$ is the mode of a primary field of spin $3/2$ with charge $+1$, i.e., from the relations of the ${\cal N}=2$ superconformal algebra
\be
[J_m, G^\pm_r] = \pm G^\pm_{m+r} \ , \qquad
[L_n, G^\pm_r] = \bigl( \frac{n}{2} - r \bigr) \, G^\pm_{m+r} \ . 
\ee
The two identities (\ref{id1}) and (\ref{id2}) are only compatible provided that 
\be\label{opposite}
(10 + 2 h_1 h_3 \psi_0) =  (4  - 4 h_1 h_3 \psi_0) \ , \qquad \hbox{i.e.} \quad h_1 h_3 \psi_0 = - \frac{N}{N+k} = - 1 \ ,
\ee
which corresponds to the case $\lambda=1$. Given the usual symmetry $\lambda\mapsto 1-\lambda$ one may wonder why $\lambda=1$ appears rather than $\lambda=0$. In fact, if one repeats the same analysis for $\bar{x}_r$ instead of $x_r$, the same analysis goes through, except that the compatibility condition is then $\lambda=0$. This is also mirrored by the fact that, in the explicit free field calculation at $\lambda=0$, the analogue of (\ref{Ue0}) vanishes, $[e_0 + \hat{e}_0, \bar{x}_{1/2}] =0$.


\bibliographystyle{JHEP}

\end{document}